\documentclass[aps, twocolumn, pre,showpacs, longbibliography]{revtex4-2}

\usepackage{amsmath,amssymb,amsthm}
\usepackage{hyperref}
\usepackage{amsthm}
\usepackage{graphicx}
\usepackage{subfigure}
\usepackage{color,bbm}
\usepackage{amsfonts}
\usepackage{mathrsfs}
\usepackage[dvipsnames]{xcolor}

\begin{document}

\date{\today}

\title{Thermodynamic Cost of Random-Time Protocols }

\author{Izaak Neri}
\affiliation{Department of Mathematics, King’s College London, Strand, London, WC2R 2LS, UK}

\begin{abstract} 

Systems  that are driven by a randomly timed,  external protocol can seemingly violate the second law of thermodynamics. We show that this thermodynamic paradox is resolved if the   outcome of the random time is stored in  a memory device.  Specifically, we show that the average work required to  erase the memory  is at  least as large as the average work gained from the protocol.   We also discuss concrete  setups that measure random times directly without continuous monitoring.     Taken together, this paper  discusses the relationship between temporal information and thermodynamics.  This framework is relevant    for external protocols employing random times, such as stochastic resetting protocols and cyclically driven heat engines that use randomly timed protocols. 
\end{abstract}

\maketitle

\section{Introduction}
In the recent literature there has been interest in thermodynamic setups for which an external agent follows a protocol that begins or ends at times that vary across different realizations of the process. Such random-time protocols can be used to convert heat into work  or to speed up  search processes.

\begin{figure}[t!]
\centering
{\includegraphics[width=0.4\textwidth]{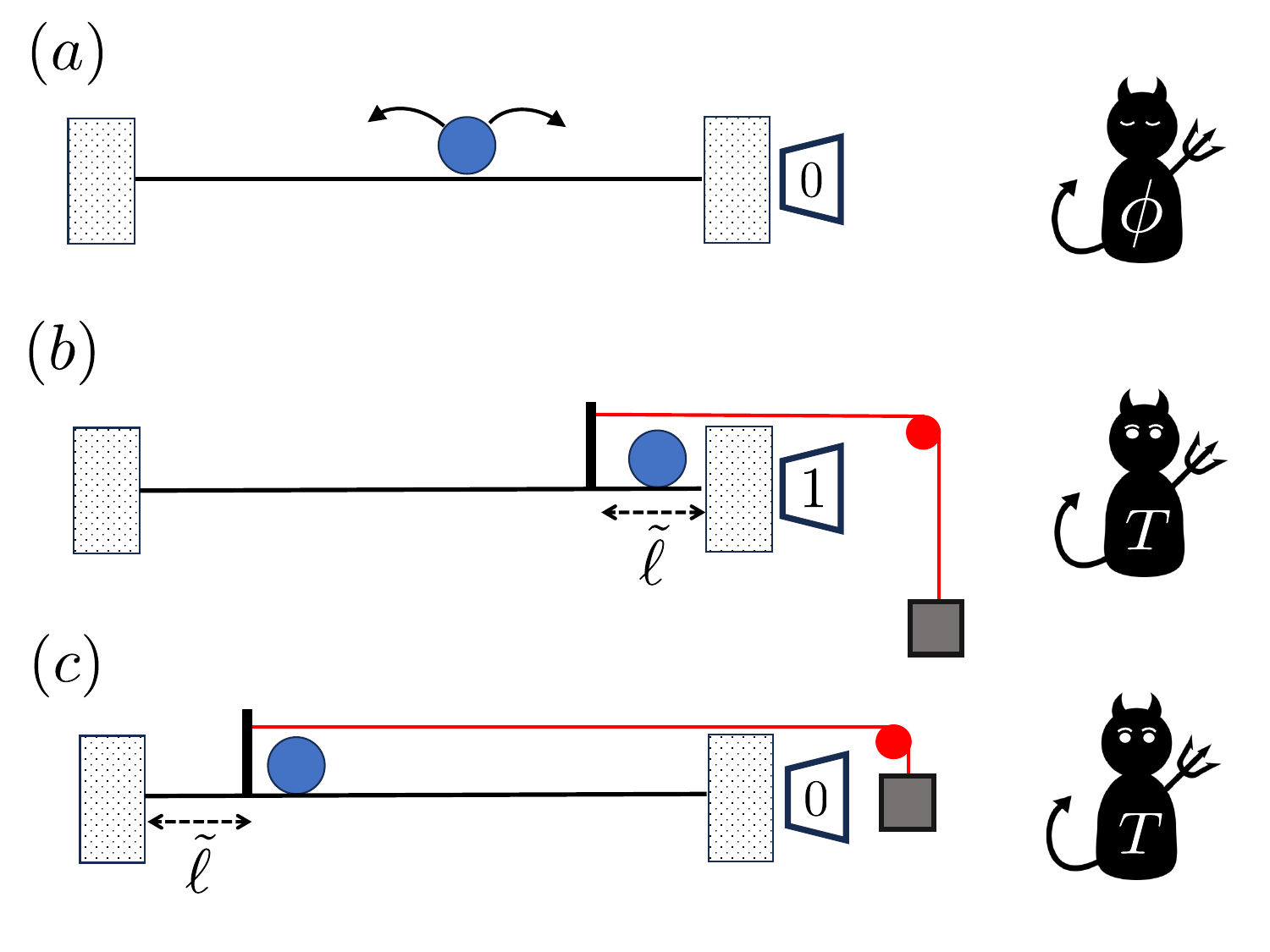}}
\caption{{\it Random-time Szilard engine}:  An example of an engine that uses temporal information to convert heat into work. Panel (a): a particle  diffuses in an interval with reflecting boundary conditions.    At one end of the segment there is a detector (depicted as a trapezoid) that is in the inactive state ($0$), a non-autonomous Maxwell demon that is dormant, and a memory device in its initial state ($\phi$).   Panel (b):   The particle has hit the right boundary.   This activates the detector (brings it into state $1$) that sends a signal to Maxwell's demon and awakens it.   The demon stores the time $T$ in its memory and initiates a protocol of fixed duration to extract work from the system.   In the present example, the demon inserts a moveable wall together with an external load at a distance $\tilde{\ell}$ from the right side.  Panel (c): The demon quasi-statically moves the wall to the left until it reaches a distance $\tilde{\ell}$ from the left boundary.   In the process, a weight is lifted and heat is converted into work.    The detector has relaxed to its inactive state $0$.   The demon  erases its memory and returns to the sleeping state in Panel~(a).}\label{fig:maxwell} 
\end{figure}

Examples of randomly timed protocols that convert heat into work, and thus seemingly realize a perpetuum mobile of the second kind, include processes that terminate upon the occurrence of a specific event~\cite{neri2019integral, neri2020second, hao}, often referred to as gambling demons~\cite{manzano2021thermodynamics, gambling2, gambling3}; cyclic protocols that incorporate a randomly timed step~\cite{ribezzi2019large, ribezzi2019workv2, garrahan2023generalized, schmitt2023information, archambault2024first}; and stochastic heat engines governed by protocols that contain a randomly time step~\cite{neri2019integral,tohme2024gambling}.   A simple example of a cyclic protocol that involves a randomly timed step and converts heat into work is shown in Fig.~\ref{fig:maxwell}. In this setup a demon inserts a moveable wall after a particle has reached the right boundary of a bounded domain. Assuming the demon can measure the time when the particle reaches the right boundary without error, the demon can convert an  arbitrary amount of  heat into work. 

 Stochastic resetting is another example of a system that is driven out of equilibrium by a randomly timed protocol.   In this case,  a  diffusing particle is returned to  its point of origin at a random time~\cite{evans2011diffusion, evans2020stochastic}. 
Stochastic resetting protocols can reduce the average time required for the particle to find a target, albeit at the expense of an energetic cost in the form of  work done on the particle.    Indeed, physically, the resetting procedure can be implemented through a resetting potential that is turned on at a random time  and  returns the particle to the origin, see Refs.~\cite{gupta2020stochastic, mercado2020intermittent,santra2021brownian,mercado2022reducing,gupta2021resetting, alston2022non,xu2022stochastic} for theoretical implementations and Refs.~\cite{reset1,goerlich2023experimental,tal2020experimental} for experiments.   A thermodynamic analysis of this  resetting setup shows that during the protocol  work is done on the particle~\cite{olsen2024thermodynamic,lahiri2024efficiency}.

In this paper, we identify a  thermodynamic cost associated with randomly timed protocols that is informational in nature.   
Specifically, drawing an analogy with Maxwell's demon in Szilard’s engine~\cite{leff1990maxwell,maruyama2009colloquium}, we argue that a non-autonomous demon implementing a random-time protocol must store the time variable in a memory device. The associated thermodynamic cost arises, as with  Szilard's engine, from restoring the memory device  to its initial state~\cite{leff1990maxwell,maruyama2009colloquium}.
As we show, apparent violations of the second law of thermodynamics are resolved once this additional cost is taken into account.

Furthermore, we introduce in this paper a non-autonomous demon that directly measures first-passage times.   Although first-passage demons have been considered before in the literature, see Refs.~\cite{ribezzi2019large,ribezzi2019workv2, garrahan2023generalized, schmitt2023information, archambault2024first},  these non-autonomous demons  measure the whole trajectory of the system up to the first-passage time through continuous monitoring.   For non-autonomous demons,  this approach  requires a 
large number of measurements, as the state of the system should be measured at regularly spaced intervals of time.     Here, instead, we develop  a simpler setup for which a non-autonomous demon directly measures the first-passage time using  a single measurement.   We  present concrete examples of setups  that  directly measure  first-passage times. 

The paper is structured as follows. In Sec.~\ref{sec:2}, we introduce the system setup for externally controlled protocols that operate with random times, focusing on the isothermal case.    In Sec.~\ref{Sec:3}, we argue, by analogy with Szilard’s engine, that a non-autonomous demon implementing a random-time protocol  requires a memory device to run the protocol.  In Sec.~\ref{sec:4}, we show that  storage of the random time is sufficient to derive the second law of thermodynamics for isothermal systems controlled by random-time protocols, further confirming that a memory device is needed in these setups.  The derivation uses  Landauer's erasure principle~\cite{landauer1961irreversibility, leff1990maxwell, maruyama2009colloquium}  and information thermodynamics    for non-autonomous Maxwell demons with a path-dependent protocol~\cite{Sag2010, Sag2012, sagawa2013information}.  The Secs.~\ref{sec:5} to \ref{sec:7} explore a paradigmatic  example of a temporal information engine, which is a random-time version of Szilard's engine, as illustrated in Fig.~\ref{fig:maxwell}.    This model is arguably the simplest example of a temporal information engine.  It is a simplification of previous setups  based on continuous Maxwell demons~\cite{ribezzi2019large, ribezzi2019workv2, garrahan2023generalized, schmitt2023information, archambault2024first}, as the demon measures the random time directly without continuous monitoring.    In Sec.~\ref{sec:5}, we consider the random-time Szilard engine  in discrete time and without measurement errors, and in Sec.~\ref{sec:6}  we consider the corresponding model in continuous time.     In Sec.~\ref{sec:7} we develop  simple models for random-time detection  without continuously monitoring the system, confirming that  first-passage time measurements are  physically sound.   In Sec.~\ref{sec:8} we show that stochastic heat engines that use random-time protocols satisfy the Carnot bound if the random time is stored in a memory.  Section~\ref{sec:8b}  compares  the setups of Maxwell demons that continuously monitor a system with those that directly measure first-passage times.   We end the paper with a discussion in Sec.~\ref{sec:9} and a few appendices that contain the  derivations of results presented in the main text.

\section{System setup: External protocols with temporal information}\label{sec:2}
 We consider  a mesoscopic system whose slow degrees of freedom are   described by a stochastic process $X(t)\in \mathcal{X}$, where   $t\geq 0$ is a discrete or continuous time index.  
 The system is subject to a potential $u(x;\lambda)$, with $x\in\mathcal{X}$ and with $\lambda$ a parameter that can be controlled  by an external agent.   Furthermore, the system is in contact with an external load that can do work on the system, or the other way around, the system can do work on the load.
 For clarity, we first focus on   isothermal systems in contact with a thermal reservoir at inverse temperature $\beta$.  In  Sec.~\ref{sec:8}  we generalize this setup to systems  interacting with two thermal reservoirs.       
 
We assume that at the initial time $t=0$  the system is  in   equilibrium  with the thermal reservoir, and thus the initial distribution of  $X(0)$  takes the form 
\begin{equation}
p_{X(0)}(x) = \exp(-\beta [u(x;\lambda_{-}) - f_{-}]  ),
\end{equation}
where   $f_{-}$ is the  free energy.   

Subsequently,  a non-autonomous demon implements a protocol $\lambda(s)$ with $s\in[0,\tau]$ that is initiated (terminated) at a time $T$  ($T+\tau$) that is dependent on the   trajectory of the process $X$.  Specifically, the  demon implements the protocol 
   \begin{equation}
  \lambda_T(t) = \left\{ \begin{array} {ccc}\lambda_{\rm -}  &{\rm if}& t\in [0,T], \\ 
  \lambda(t-T) &{\rm if}& t\in [T,T+\tau], \\ 
  \lambda_{+} &{\rm if }& t \geq T+\tau.
  \end{array}\right.  \label{eq:protocol}
  \end{equation} 
  Note that  in  standard setups in information thermodynamics the demon  chooses  one protocol out of   two or more  options based on the measurement of $X$ at a fixed time~\cite{szilard1929entropieverminderung, leff1990maxwell, horowitz2011thermodynamic}.      In contrast, here the  protocol  is always the same, but the time $T$ when the protocol  starts differs from one realization to the other.

The random time $T$ is  a function of the trajectory $X^{\infty}_0$; we use the notation $X^{t}_s = \left\{X(u): u\in [s,t]\right\}$ for trajectories.   Due to 
causality,  we assume that $T$ is statistically independent of $X^{\infty}_T$ (i.e., $T$ is a stopping time ~\cite{bremaud2001markov, rogers2000diffusions, neri2019integral}).

The work done by the external load on the system is given by~\cite{sekimoto1998langevin,seifert2012stochastic} 
\begin{equation}
W(t) = \int^t_0 \left.\frac{\partial u(X(s);\lambda)}{\partial \lambda}\right|_{\lambda=\lambda_T(s)} \frac{{\rm d}\lambda_T(s)}{{\rm d}s} {\rm d}s,
\end{equation}
with $W(t)>0$ when the load does work on the system and $W(t)<0$ when the system does work on the load.   By the first law of thermodynamics, the heat exchanged with the thermal reservoir is given by  
\begin{equation}
Q(t) = W(t) - u(X(t),\lambda_T(t)) + u(X(0),\lambda_T(0)),
\end{equation}
with $Q(t)$ the heat released into the reservoir.

In what follows, we denote averages over multiple realisations of the process $X$ by $\langle \cdot\rangle$.

\section{Memory storage  of the random time $T$ } \label{Sec:3}
We argue that a non-autonomous Maxwell demon needs to store the random time $T$ in a memory to run the protocol $\lambda_T$.  The underlying reason is the same as why Maxwell’s demon in Szilard’s engine requires a memory~\cite{bennett1982thermodynamics, leff1990maxwell, maruyama2009colloquium}: if an external protocol depends on a random variable,  then this random variable must be stored in a memory, as otherwise it is not possible for the demon to evaluate the protocol.     The only conceptual difference between   Maxwell's demon in Szilard's engine and non-autonomous demons in  random-time protocols is that in the former the random variable is a spatial coordinate, while in the latter it is a temporal coordinate.   Since there is no justification for treating time variables differently from spatial variables, we require that  demons also  need to store random times $T$ in a memory.

Let us slightly elaborate on the above argument that makes an analogy between protocols that depend on spatial information and those that depend on temporal information.      In Szilard's engine, a  non-autonomous  Maxwell demon measures  at a fixed time, say $t=\tau_{\rm m}$, the position $X(\tau_{\rm m})$  of a particle diffusing in a bounded domain~\cite{szilard1929entropieverminderung, leff1990maxwell, maruyama2009colloquium}.      As in Szilard's engine the purpose of the measurement is to determine whether the   particle is located to the left or right of the center, we consider the coarser observable  $Y\in \left\{{\rm R}, {\rm L}\right\}$, with $Y={\rm R}$ if $X(\tau_{\rm m})$ is to the right of the center, and $Y={\rm L}$ if $X(\tau_{\rm m})$ is to the left of the center.    The corresponding protocol takes the general form 
\begin{equation}
\lambda_{Y}(t) =  \left\{ \begin{array} {ccc}\lambda_{\rm -}  &{\rm for}& t\in [0,\tau_{\rm m}], \\ 
  \lambda_{\rm L}(t) &{\rm for}& t\in [\tau_{\rm m},\tau+\tau_{\rm m}] \  {\rm if} \  Y  = {\rm L}, \\ 
  \lambda_{\rm R}(t)&{\rm for }& t\in [\tau_{\rm m},\tau+\tau_{\rm m}] \  {\rm if} \  Y  = {\rm R}. \end{array}\right. \label{eq:standard}
\end{equation}

Comparing the protocols (\ref{eq:standard}) and (\ref{eq:protocol}), we note several  similarities. Notably, in both cases the demon must know the outcome of a random variable to evaluate the protocol: in (\ref{eq:protocol}), the demon needs to know $T$ to evaluate $\lambda_T(t)$ for times $t>T$, while in (\ref{eq:standard}), the demon needs to know $Y$ to evaluate $\lambda_Y(t)$ for $t>\tau_{\rm m}$. In the latter case, it has been argued that the demon requires a memory to store the outcome of $Y$ in order to evaluate the protocol~\cite{szilard1929entropieverminderung, leff1990maxwell, maruyama2009colloquium}. Since there is no apparent reason to treat temporal information differently from spatial information, we conclude that $T$ should likewise be stored in a memory.

\section{Second law of thermodynamics for non-autonomous demons that use random times} \label{sec:4}
Using an analogy with Szilard's engine, we have argued that spatial and temporal information should be treated on equal footing, and thus the outcome of $T$ must be stored in a memory.   The question is now whether this is sufficient to uphold the second law of thermodynamics.
In this section, we show that this is indeed the case:  non-autonomous demons that use  protocols with random times satisfy the second law of thermodynamics, provided that the random time $T$ is stored in a memory device.

For  cyclic processes in isothermal systems,   the Kelvin version of the second law of thermodynamics states that~\cite{fermi} 
\begin{equation}
\langle W_{\rm useful}\rangle \geq 0, \label{eq:wusefulSecond}
\end{equation}
where $\langle W_{\rm useful}\rangle$ is the  average  work  that has been done on the system.    Equation (\ref{eq:wusefulSecond}) states that the useful work cannot be on average negative, or in other words, it is not possible to absorb heat from a thermal reservoir and convert it into useful work.  

To determine the useful work, let us first consider the work done by  the external load on the system.  Applying stochastic thermodynamics arguments akin to those in the  works of Sagawa and Ueda~\cite{Sag2010, Sag2012, sagawa2013information} to setups with randomly timed protocols, we derive in the  Appendix~\ref{app:A} the inequality 
 \begin{equation}
\beta(\langle W\rangle-\Delta f)  \geq   -   H(T)  ,   \label{eq:WBound2}
\end{equation}
where  $\langle W\rangle = \langle W(T+\tau)\rangle $ is the total work done  by the load on the system, $\Delta f = f_+-f_-$ is the free energy difference between the end state  ($\lambda=\lambda_+$)  and the initial state  ($\lambda=\lambda_-$),  and where 
\begin{equation}
H(T) = - \sum_{t\geq 0}p_T(t)\ln p_T(t) \label{eq:entropyDef}
\end{equation}
is the Shannon entropy of  the stopping time $T$ measured in nats;  here $p_T$ is the probability distribution of $T$.       The inequality (\ref{eq:WBound2}) states that random-time protocols can thermodynamically perform work on an external load, i.e., $\langle W\rangle<0$. For cyclic protocols with $\Delta f = 0$ the work is bounded by 
 \begin{equation}
 \langle W\rangle \geq-   H(T).
 \end{equation}
Thus, the larger  the information contained in the random time $T$, the more work can, in principle, be extracted from the thermal reservoir.

We note that the Eq.~(\ref{eq:WBound2}) is a stochastic thermodynamics result.   It can be derived in the setup of Markov processes that satisfy  local detailed balance  (see  Appendix~\ref{app:A}).  An alternative derivation of the  inequality (\ref{eq:WBound2}) that is based on continuously monitoring the system  is presented in Ref.~\cite{schmitt2023information}. This is because at the level of mathematical formulation, both the sleeping demon and the conventional continuous demon can be described in the unified setup of an error-free non-autonomous Maxwell demon with a path-dependent protocol $\lambda(t;X^t_0)$~\cite{Sag2010, Sag2012, sagawa2013information}.  
Furthermore, following Ref.~\cite{ashida2014general}, the inequality (\ref{eq:WBound2}) can be further refined by taking into account the unavailable information. However, our purpose is to derive the second law of thermodynamics for randomly timed protocols, and for this purpose the inequality (\ref{eq:WBound2}) provides the most parsimonious explanation. 

To derive the Kelvin statement of the second law, (\ref{eq:wusefulSecond}), we further consider that the 
demon needs a memory to  store the random time $T$.  If the process is  cyclic, then the memory  should also be restored to its initial state.       According to Landauer's erasure principle, the average work  $\langle W_{\rm erase} \rangle$ needed   to erase $H(T)$ nats of information   is lower bounded by~\cite{landauer1961irreversibility, leff1990maxwell, maruyama2009colloquium} 
 \begin{equation}
 \beta \langle W_{\rm erase}\rangle \geq  H(T)   = - \sum_{t\geq 0}p_T(t)\ln p_T(t).  \label{eq:erasure} 
 \end{equation}   

Combining the information theoretical bound (\ref{eq:WBound2}) with Landauer's erasure principle (\ref{eq:erasure}), we find 
\begin{equation}
\langle W_{\rm useful}\rangle = \langle W\rangle + \langle W_{\rm erase}\rangle  \geq 0, \label{eq:wuseful}
\end{equation}
and thus we have recovered the second law of thermodynamics.

\section{Random-time Szilard engine: discrete version}\label{sec:5}
To clearly illustrate the  concepts of temporal information dynamics, we analyse  an information engine that uses temporal information  to convert heat into work.   This  temporal information engine is  a 
random-time version of Szilard's  engine~\cite{szilard1929entropieverminderung, leff1990maxwell} (see Fig.~\ref{fig:maxwell}).

\begin{figure}[t!]
\centering
{\vspace{-4mm}\includegraphics[width=0.4\textwidth]{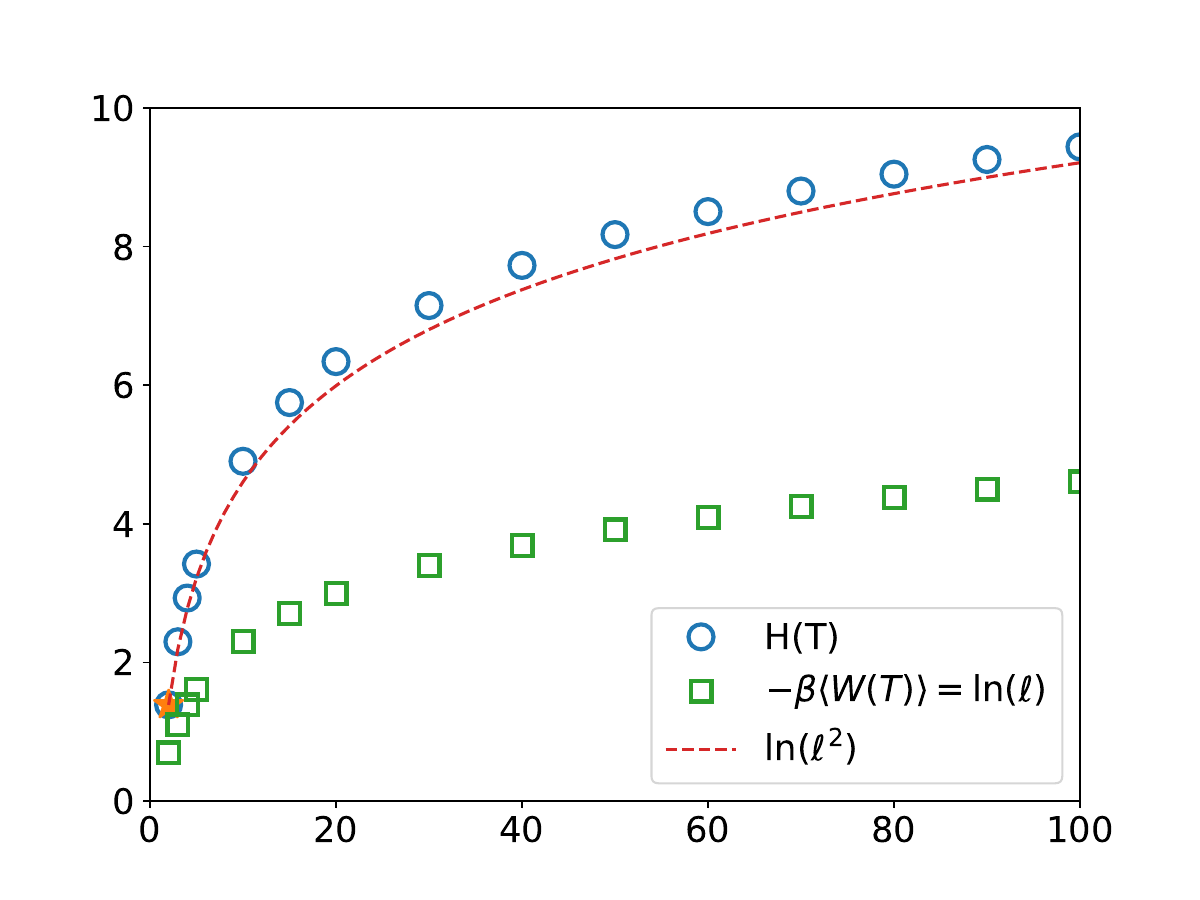}}
 \put(-100,-2){$\ell$}
\caption{{\it   Random-time Szilard engine  in discrete time without measurement error} (see Sec.~\ref{sec:5}).  Comparison between the work $-\beta\langle W\rangle=\ln(\ell) $  (green squares) done by the Szilard engine on the external load  and the entropy $H(T)$ (blue circles) of the  stored time $T$   as a function of the length $\ell$ of the lattice; the hopping probability  
$p_0=1$.    The blue circles denote the entropy  $H(T)$  of   the probability mass function of  $T={\rm min}\left\{t\geq 0: X(t)=\ell\right\}$  obtained from repeated simulations of the random walk process with a uniform initial distribution for $X(0)$.   
 The (orange) star denotes the value  $H(T) =2 \ln 2$ for $\ell=2$ (see Appendix~\ref{App:B}), and the red dotted line plots the function $\ln \ell^2$.     
}\label{fig:plotEntropy} 
\end{figure}

  The random-time Szilard engine consists of a particle diffusing in a box,  a random-time detector, a sleeping demon, and a memory device.   The sleeping demon is awoken by the detector at the first moment when  the particle  collides  with the right boundary.    At that moment, the demon  inserts a moveable wall near the right boundary,  attaches a load and a pulley on the right side of the moveable wall, and quasi-statically moves the wall to the left boundary.     Importantly,   the information that the  demon gains  in this protocol is purely time-based.   This is different from Maxwell's demon in the Szilard engine, which places  the load on the right or left side of the moveable wall, depending on the outcome of the measurement~\cite{szilard1929entropieverminderung, leff1990maxwell}.  
  
  Note that sleeping demons simplify setups based on continuous Maxwell demons.  Continuous Maxwell demons  measure the full trajectory $X^T_0$ through continuous monitoring~\cite{ribezzi2019large,ribezzi2019workv2, garrahan2023generalized, schmitt2023information}.  Instead,   the sleeping demon  directly measures $T$ without continuous monitoring.  This is possible within classical physics as we show in    Sec.~\ref{sec:7}.

In  discrete time, the  working substance is a  particle that performs an unbiased  random walk on a one-dimensional lattice of length $\ell$ with reflecting boundary conditions.     The position $X(t)$ of the particle at time $t\in \mathbb{N}$  takes values $X(t)\in \left\{1,2,\ldots,\ell \right\}$.    At every time step, the particle  moves to one of its nearest neighbours with   probability $p_0/2$ (two neighbours for bulk sites, $X(t)\notin  \left\{1,\ell\right\}$, and one neighbour for  boundary sites, $X(t)\in \left\{1,\ell\right\}$), and  otherwise the particle does not move.    Note that the particle diffuses because it is in contact with a thermal reservoir at inverse temperature $\beta$.  

We assume that the random-time detector is error-free, so that the sleeping demon receives a signal from the detector the first time $T={\rm min}\left\{t\geq 0: X(t)=\ell\right\}$ when the particle reaches the last site of the lattice, at which point the demon awakens.  The demon stores  the time $T$ in its memory, inserts a moveable wall between the lattice sites $\ell-1$ and $\ell$, and attaches a load to a pulley on the right side of this wall.      The demon quasi-statically moves the wall  to the left until it reaches the left-hand side  of the lattice at   time $T+\tau$, with $\tau$ significantly larger than the relaxation time to equilibrium.  At time $T+\tau$, the demon removes the wall,   resets its memory to its initial state, and  goes back to sleep.  

If the demon moves the wall quasi-statically, then the average work  done by the particle on the load  is given by 
$-\beta \langle W(T+\tau)\rangle =  \ln \ell $.   Indeed, 
 by the first law of thermodynamics, the work performed on the load equals the heat exchanged with the thermal environment, and since the protocol is quasi-statically slow, the heat exchanged equals  the  entropy difference times the  temperature.  Thus we can conclude that the average work grows logarithmically with  $\ell$~\cite{leff1990maxwell}.    On the other hand, the entropy $H(T)\approx \ln (\ell^2)$, as time $T\sim \ell^2$.  Thus, the average work the temporal information engine exerts on the load is smaller  than the average work needed to restore the memory to its original state, in correspondence with Eqs.~(\ref{eq:WBound2}) and (\ref{eq:wuseful}).   In Fig.~\ref{fig:plotEntropy} we verify that indeed  $-\beta \langle W(T+\tau)\rangle/H(T)<1$ for all lattice lengths $\ell$.

\section{Random-time Szilard engine: continuous version}\label{sec:6}
We consider a continuous version of the random-time Szilard engine.   In this case, the working substance is a Brownian particle  with diffusion constant $D$ that diffuses in a one-dimensional,  bounded domain of length $\ell$. % so that ${\rm d}X/{\rm d}t = \sqrt{2D} {\rm d}W/{\rm d}t$ with   $W(t)$ a standard Wiener process.  
This model is not only physically more realistic than its discrete version, it is also analytically solvable.   

\begin{figure}[]
\centering
{\includegraphics[width=0.35\textwidth]{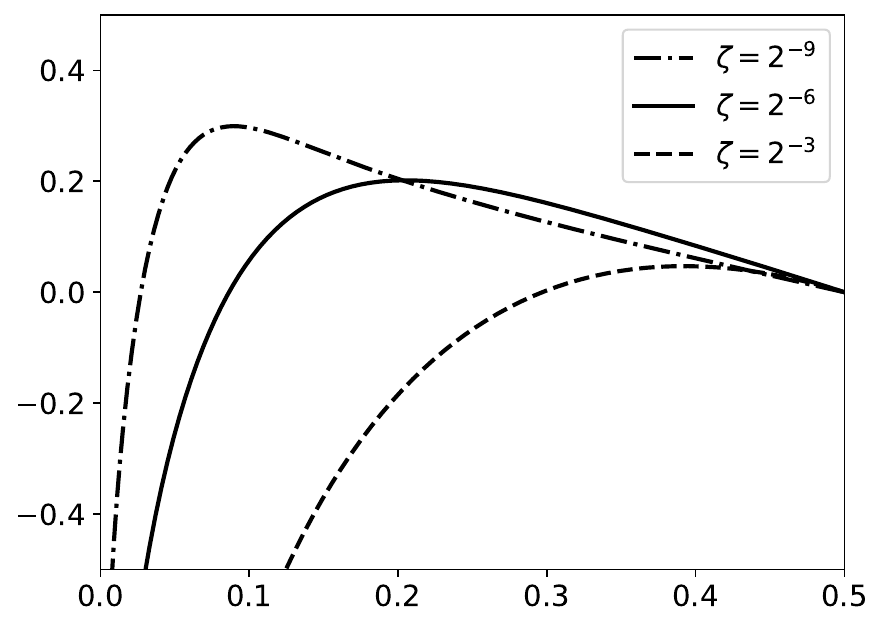}} \put(-212,65){$-\frac{ \beta\langle W \rangle}{H(T)}$}
 \put(-92,-7){$\tilde{\ell}/\ell$}
\caption{{\it Random-time Szilard engine in continuous time without measurement error} (see Sec.~\ref{sec:6}). The ratio between the  work $-\beta\langle W\rangle$ done by the system on the  external load  and the entropy $H(T)$  as a function of 
the length ratio $\tilde{\ell}/\ell$ for three given values of $\zeta = D\theta/\ell^2$.  See Appendix~\ref{app:C}   for explicit expressions of the plotted functions.      }\label{fig:work} 
\end{figure}

In continuous time, the entropy  $H(T)=+\infty$, and hence the demon needs an infinite amount of memory space to store $T$.  Correspondingly, the demon can exchange an infinite amount of heat into work, as the diffusive particle can be localised arbitrary close near the right boundary~\cite{ribezzi2019large}. 

To make the model physically realistic, we  assume that the demon has a finite memory device at its disposal.   Consequently, the random time $T$ when the moveable wall is inserted into the system, and which is stored in the memory device, is different from the first-passage time 
\begin{equation}
T_{\rm fp} = {\rm min}\left\{t\geq 0: X(t)=\ell\right\}. \label{eq:Tfp}
\end{equation}
For the sake of simplicity, we consider a discretisation of the form,  
\begin{equation}
T = {\rm min}\left\{m \:\theta:m\in \mathbb{N} \ {\rm and} \ m\theta\geq T_{\rm fp}\right\},  \label{eq:TContDef}
\end{equation}
 where $\theta>0$ is a time interval.  The entropy  $H(T)$  is a function of the ratio 
 \begin{equation}
 \zeta=\frac{\theta D}{\ell^2} \label{eq:zetaDef}
\end{equation}
 between the  average squared  distance $\theta D$ travelled by the Brownian particle in a time-interval $\theta$ and the squared length $\ell^2$ of the domain. For small values of $\zeta$  the entropy is well approximated by $-\ln \zeta$ (see Appendix~\ref{app:C}).  
 
Discretisation of time leads to loss of temporal information, and according to the inequality (\ref{eq:WBound2}),  a corresponding loss in the amount of work that an information engine can do on an external load.   Indeed, 
 since $T$ is different from $T_{\rm fp}$, it holds in general that $X(T) \neq\ell$.   To assure that the particle is located to the right of the moveable wall, the sleeping demon inserts the wall at a distance $\tilde{\ell}$ from the right boundary of the domain. Analogously, the wall is removed at a distance $\tilde{\ell}$ from the left boundary of the domain.     The average work takes thus the form 
 \begin{equation}
\beta\langle W(T+\tau)\rangle =  p_+ \ln \frac{\tilde{\ell}}{\ell-\tilde{\ell}} + (1-p_{+})  \ln \frac{\ell-\tilde{\ell}}{\tilde{\ell}}, \label{eq:Waverage}
\end{equation}
where  
\begin{equation}
p_+ = {\rm Prob}(X(T)\geq \ell-\tilde{\ell}) \label{eq:pPDef}
\end{equation}
is the probability  that the particle is  located at the right of the inserted wall.    The probability $p_+$ is a function of the ratios $\zeta$ and $\tilde{\ell}/\ell$, with $p_+=0$ ($p_+=1$) if $\tilde{\ell}=0$ ($\tilde{\ell}=\ell$), as we also show in Appendix~\ref{app:C}.     

In Fig.~\ref{fig:work} we plot the ratio  between the average extracted work $-\beta \langle W\rangle$ and the amount of temporal information $H(T)$ as a function of the ratio $\tilde{\ell}/\ell$ for three values of $\zeta$.   The ratio is smaller than $1$, in correspondence with the bound~(\ref{eq:WBound2}), expressing a trade off between temporal information and extracted work.   The average work reaches a maximum at an intermediate value of  $\tilde{\ell}/\ell\in (0,1)$, corresponding with the optimal placement  of the inserted wall for a given value of $\zeta$ (see Appendix~\ref{app:C} for a more detailed analysis).

\section{Random-time detectors}\label{sec:7}
The random-time Szilard engine of Fig.~\ref{fig:maxwell} contains a random-time detector, which so far we have assumed to be perfect.   We develop now  physical models for random-time detectors.  We first consider the case of a   discrete random-time Szilard engine.   

We assume the  state of the detector, $Y(t)$, takes two possible values, inactive $Y(t) = \downarrow$ and active $Y(t)=\uparrow$.    These two levels have  energies $\epsilon_\downarrow=0$ and $\epsilon_\uparrow>0$, respectively.  We assume that the  detector is  in    equilibrium with the thermal reservoir at inverse temperature $\beta$, just like the  diffusing particle.      
The sleeping demon measures the transitions from $\uparrow$ to the ground state $\downarrow$, i.e., 
\begin{equation}
T = {\rm min}\left\{t\geq 0:Y(t-1)=\uparrow, \: Y(t) = \downarrow  \right\}.
\end{equation}
For instance, we can imagine that during the  transition  a photon of a specific frequency is emitted,  and the demon is highly responsive to this frequency. 

\begin{figure}[t!]
\centering
{\includegraphics[width=0.4\textwidth]{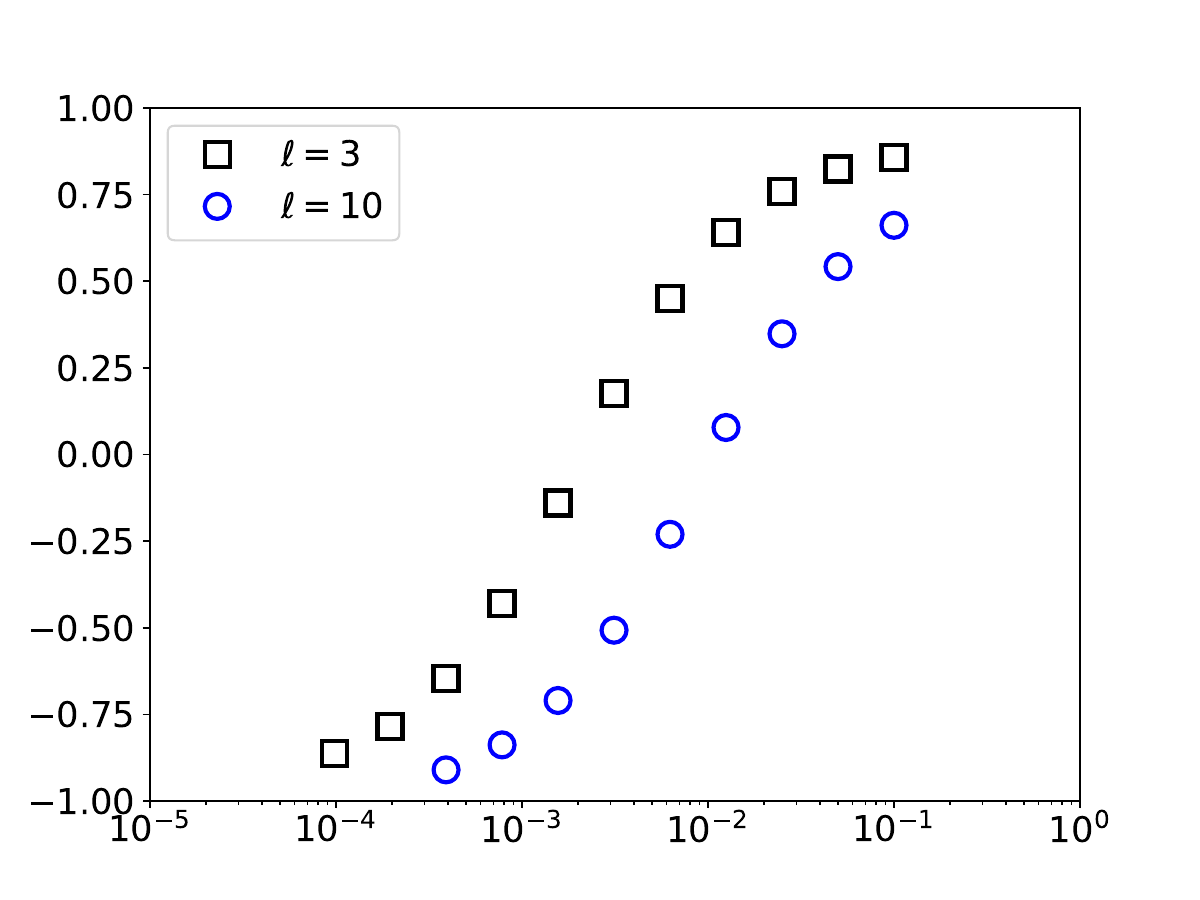}}
 \put(-235,70){$-\frac{\beta  \langle W \rangle}{\ln(\ell-1)}$}
 \put(-110,-2){$\pi_\ell/p_0$}
\caption{{\it  Random-time Szilard engine   with measurement error} (see Sec.~\ref{sec:7}). Average work  that a random-time Szilard engine exerts on an external load as a function of the parameter $\pi_\ell/p_0$ that characterises the quality of the detector.   The  other detector parameters are   $\pi_0=0.0001$ and  $\beta \epsilon_\uparrow = 4 $.   The hopping probability 
 $p_0=0.1$  and $\ell$ is as given in the legend.  The average work is obtained from  the formula $-\beta \langle W\rangle = (2p_+-1)  \ln (\ell-1)$, where  $p_+$ is the probability that $X(T)=\ell$; this is Eq.~(\ref{eq:Waverage}) with $\tilde{\ell}=1$.     Each marker is an average over $1e+6$ simulated samples.  }\label{fig:plotDet} 
\end{figure}

The transition probabilities  between the two detector  states are assumed to take the form  $
\pi_{\uparrow} = \pi_0 (1-\delta_{X(t),\ell})+ \pi_{\ell}\delta_{X(t),\ell}$ 
and 
$\pi_{\downarrow} = [\pi_0 (1-\delta_{X(t),\ell}) + \pi_{\ell}\delta_{X(t),\ell}]e^{\beta \epsilon_\uparrow}$, for transitions towards the $\uparrow$ and $\downarrow$ states, respectively.  Here, 
 $\delta$ denotes the Kronecker delta function, and $\pi_0,\pi_\ell\in [0,\exp(-\beta \epsilon_\uparrow)]$. Note that the detector is coupled with the particle's state as the transition probabilities $\pi_\downarrow$ and $\pi_\uparrow$ depend on $X(t)$.

An effective detector requires tuning $\pi_0$, $\pi_\ell$, and $\epsilon_\uparrow$ to meet three conditions.  First, we  set $\exp(-\beta \epsilon_\uparrow) \ll  1/\ell$, so that the detector occupies mostly the ground state $\downarrow$.    Second, we set   $\pi_0 \ll p_0/\ell^2$, so that the detector is unlikely to activate when the particle is not on the $\ell$-th site. Third, we set $\pi_\ell \gg p_0$ so that the detector activates as soon as the particle reaches the  $\ell$-th site.  
In Fig.~\ref{fig:plotDet} we confirm that under these conditions the detector functions well, as,  similar to the case of a perfect detector, $\beta\langle W(T+\tau)\rangle \approx -\ln (\ell-1)$.      In fact, we observe that the random-time detector works already well for relatively small values of $\pi_\ell \gtrsim 0.1 p_0$.

\begin{figure*}[]
\centering
{\includegraphics[width=0.5\textwidth]{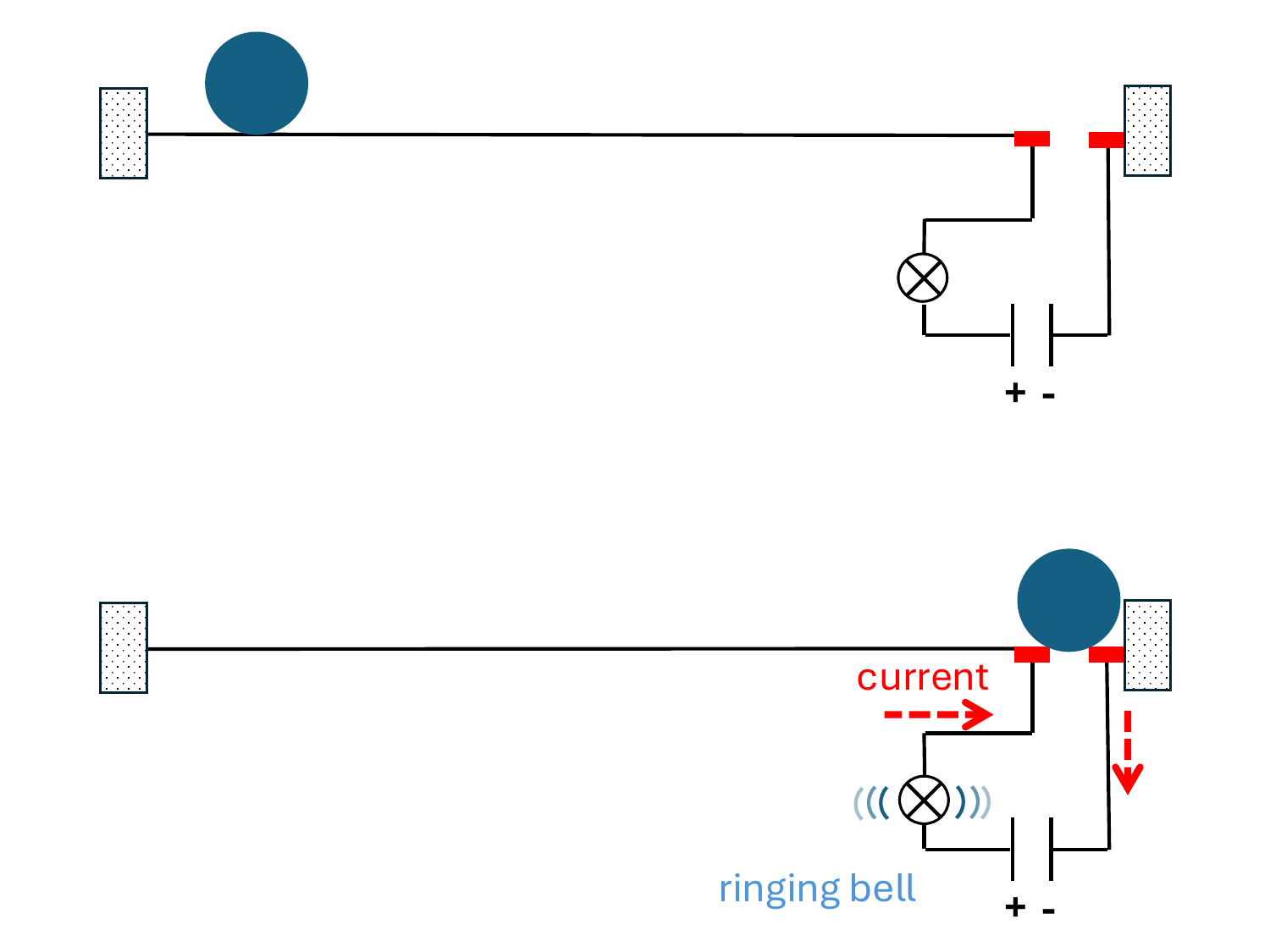}} 
\caption{ Example of a setup that   measures the first time when a particle reaches the end of a bounded domain without continuously monitoring the system.   At the right end of the segment there is an electrical  circuit that is usually  turned off (see top panel).  The diffusing colloidal particle is made of a material of high conductance.     When the particle reaches the right end, it fills the gap between two electrodes and the electrical circuit is switched on (see bottom panel).  Once the circuit is switched on, a bell rings that alerts the external observer (i.e., it "awakens" the sleeping demon).   }\label{fig:exampleMeas} 
\end{figure*}

An alternative,  possibly more accessible, way of measuring a first-passage time  is illustrated in Fig.~\ref{fig:exampleMeas}.   In this case, a diffusing particle that is made of a material that conducts electricity well  is immersed into an insulating and  viscous environment.   When the particle reaches the end of the segment, it   switches on an electrical circuit, and the current in this circuit activates an alarm that awakens the sleeping demon.   

Note that the function of the detector in both of the above examples is to awaken the demon.
Once the demon is alerted, it measures the time 
$T$,  stores $T$ in a memory device, and initiates the protocol $\lambda_T(s)$.  One caveat here is that the demon needs a clock  to measure time.  We elaborate on this in the discussion section of the paper.

\section{Carnot efficiency for   cyclic heat engines that use randomly timed protocols}\label{sec:8}

 We evaluate the efficiency of stochastic heat engines that rely on temporal information, see for example the  model of Ref.~\cite{tohme2024gambling}.   Heat engines are systems in   contact with two  or more thermal reservoirs at different temperatures.  
Unlike isothermal systems,  cyclic processes coupled to two reservoirs at different temperatures can perform  average  work on an external load. However, the second law of thermodynamics limits the relative amount of useful work that can be done on an external load.  For cyclic processes in contact with two thermal reservoirs, the average efficiency of a stochastic heat engine is bounded from above by the Carnot bound
\begin{equation}
\eta  = \frac{\langle W_{\rm useful}\rangle }{\langle Q_{\rm h}\rangle}\leq 1-\frac{\beta_{\rm h}}{\beta_{\rm c}},  \label{eq:carnot}
\end{equation} 
where $Q_{\rm h}$ is the heat released into the hot reservoir, $\beta_{\rm h}$ is the inverse temperature of the hot reservoir, and $\beta_{\rm c}>\beta_{\rm h}$ is the inverse temperature of the cold reservoir; note that in the typical operation of a heat engine, both   $\langle W_{\rm useful}\rangle$ and $\langle Q_{\rm h}\rangle$ are negative.  

Interestingly, for protocols that use random times, the efficiency can (apparently) exceed the Carnot bound, as shown in the recent paper~\cite{tohme2024gambling}.  Nevertheless, we demonstrate here that the Carnot bound is recovered when accounting for the cost associated with storing the random time $T$.

\subsection{Information theoretical bound}
We first derive the equivalent of Eq.~(\ref{eq:WBound2}) for heat engines.    Specifically, if we define the efficiency 
\begin{equation}
\eta_{\rm G} = \frac{\langle W\rangle}{ \langle Q_{\rm h}\rangle},  
\end{equation}
where $\langle W\rangle$ is the average work done by an external load on the system, then 
\begin{equation}
\eta_{\rm G}\leq  \left(1-\frac{\beta_{\rm h}}{\beta_{\rm c}}\right)  - \frac{H(T)}{\beta_{\rm c}\langle Q_{\rm h}\rangle}.\label{eq:TCarnot}
\end{equation}
Note  the right-hand side of (\ref{eq:TCarnot}) contains an information theoretical term corresponding with the additional work that we  can extract by using the information contained in $T$.    This term is positive in the normal operation of a heat engine ($\langle Q_{\rm h}\rangle<0$) and thus super-Carnot efficiencies are in principle allowed when $H(T)$ is large enough.  

To derive  (\ref{eq:TCarnot}), note that Eq.~(\ref{eq:WBound2})  generalises into   
  \begin{equation}
  \langle S_{\rm tot}(T+\tau)\rangle \geq  -H(T),
  \end{equation}
  where $\langle S_{\rm tot}(T+\tau)\rangle$ is the average entropy production.   For a cyclic process in contact with two thermal reservoirs, we obtain  
\begin{equation}
\beta_{\rm c} \langle Q_{\rm c}\rangle + \beta_{\rm h} \langle Q_{\rm h}\rangle \geq -H(T). \label{eq:boundC}
\end{equation}
Here, $\langle Q_{\rm c}\rangle = \langle Q_{\rm c}(T+\tau)\rangle$ and $\langle Q_{\rm h}\rangle = \langle Q_{\rm h}(T+\tau)\rangle$ quantify  the heat released to the      cold and the hot thermal reservoirs, respectively, after the completion of one cycle.      Note that the cycle terminates at a random time $T+\tau$, as there is one step in the protocol that initiates at a random time $T$.  

The  first law of thermodynamics for this engine reads
\begin{equation}
\langle Q_{\rm c}\rangle   = \langle W\rangle  - \langle Q_{\rm h}\rangle,  \label{eq:firstx}
\end{equation}
where we have used the following sign conventions: $\langle W\rangle>0$ implies that the external load does work on   the system, and $\langle Q_{\rm c}\rangle>0$ (or $\langle Q_{\rm h}\rangle>0$) implies that heat flows from the system to the thermal reservoir.    
Using the first law (\ref{eq:firstx}) in Eq.~(\ref{eq:boundC}) to eliminate $\langle Q_{\rm c}\rangle$, we get that 
\begin{equation}
- \langle W\rangle  \leq  -\left(1-\frac{\beta_{\rm h}}{\beta_{\rm c}}\right) \langle Q_{\rm h}\rangle + \frac{H(T)}{\beta_{\rm c}}. \label{eq:bound2}
\end{equation}
Assuming that  $-\langle Q_{\rm h}\rangle > 0$, i.e., the system takes heat from the  hot thermal reservoir, we get that 
\begin{equation}
\frac{ \langle W\rangle}{\langle Q_{\rm h}\rangle}  \leq  \left(1-\frac{\beta_{\rm h}}{\beta_{\rm c}}\right)  - \frac{H(T)}{\beta_{\rm c}\langle Q_{\rm h}\rangle},\label{eq:bound3}
\end{equation}
which we recognise as the inequality (\ref{eq:TCarnot}) that we were meant to derive.

  \subsection{Analysis  with memory storage: recovering the Carnot bound }
As argued before,  a non-autonomous demon needs a memory device to store  the outcome of the random time $T$.   By Landauer's erasure principle,  storage of $T$ comes with an additional cost, and here we show that this cost is sufficient to recover the Carnot bound (\ref{eq:carnot}).      

Using, as in Eq.~(\ref{eq:wuseful}) that the useful work is 
\begin{equation}
\langle W_{\rm useful}\rangle = \langle W\rangle +  \langle W_{\rm erase}\rangle, 
\end{equation}
 where $ \langle W_{\rm erase}\rangle$ is the average work required to erase the memory,
we find for the   efficiency of the Carnot heat engine 
 \begin{equation}
 \eta =  \frac{\langle W\rangle +  \langle W_{\rm erase}\rangle}{\langle Q_{\rm h}\rangle} = \eta_{\rm G} + \frac{\langle W_{\rm erase}\rangle}{\langle Q_{\rm h}\rangle}. \label{eq:eta2}  
 \end{equation}
In typical situations  the signs  are  such that   $\langle W\rangle<0$  (the system performs work on an external load) and $ \langle W_{\rm erase}\rangle>0$ (the demon (or system) does work on the memory device).   

Using  the information theoretical bound (\ref{eq:TCarnot}) in (\ref{eq:eta2}),  we find 
 \begin{equation}
 \eta \leq  \left(1-\frac{\beta_{\rm h}}{\beta_{\rm c}}\right)  - \frac{H(T)-\beta_{\rm c}\langle W_{\rm erase}\rangle}{\beta_{\rm c}\langle Q_{\rm h}\rangle} . \label{eq:eta3}
 \end{equation}

By Landauer's  erasure principle, the work required to erase the  memory device is bounded from below by: 
  \begin{equation}
  \langle W_{\rm erase}\rangle \geq \frac{ H(T)}{\beta_{\rm c}}.  \label{eq:WERase}
 \end{equation}
Here, we  have chosen to release the heat from the memory device  into the colder reservoir, as   $1/\beta_{\rm c}\leq 1/\beta_{\rm h}$.    
 
 Using  (\ref{eq:WERase}) in (\ref{eq:eta3}) and the inequality $\langle Q_{\rm h}\rangle<0$,  we recover  Carnot's bound (\ref{eq:carnot}). 
In conclusion, a  cyclic heat engine   satisfies the Carnot bound, even if it uses a protocol with a random time $T$.

\section{Measuring first-passage times: comparison between continuous monitoring and  direct methods  }\label{sec:8b}
The sleeping demons in the present paper measure first-passage times directly.   This approach is different from   first-passage time measurements considered in the literature that measure first-passage times  through continuous monitoring~\cite{ribezzi2019large, ribezzi2019workv2, garrahan2023generalized, schmitt2023information, archambault2024first}.     We discuss here how in these two approaches  the cost of  measurement scales   with the extracted work.

\subsection{Continuous monitoring}
Let us first revisit the continuous monitoring formalism, see Refs.~\cite{ribezzi2019large, ribezzi2019workv2, garrahan2023generalized, schmitt2023information, archambault2024first} and ~\cite{Sag2012, ashida2014general, potts2018detailed}.  In the continuous monitoring framework,  the demon measures $X$ at regularly spaced time intervals $0,\delta t, 2\delta t,\ldots$.  The demon inserts a moveable wall at position $\ell-\tilde{\ell}$   the first time  when  $X(n\delta t)\in [\ell-
\tilde{\ell},\ell]$  (notice that  it is more efficient to measure a binary variable that indicates whether  $X(n\delta t) \in [\ell-
\tilde{\ell},\ell]$).   To avoid errors, we need to consider the limit of $\delta t\rightarrow 0$ which is the limit of continuous monitoring.  

In continuous monitoring, the cost of measurement scales with the system size and with the work extracted.  Indeed, let us denote by  $\delta s_{\rm meas}>0$  the  average amount of  entropy produced in a single measurement.       
Then, the  average amount of dissipation due to measurements in continuous Maxwell demons equals, on average,
\begin{equation}
\langle S_{\rm meas} \rangle = \delta s_{\rm meas} \frac{\langle T\rangle}{\delta t}. \label{sec:meas}
\end{equation}
In the limit of $\delta t\rightarrow 0$, the measurement cost gets thus large, unless we can compensate with a small cost $\delta s_{\rm meas}$ for a single measurement.  More troublesome is that  the cost of measurement increases with the size of the domain, as $\langle T\rangle\sim \ell^2$.      This implies that the cost of measurement grows faster with $\ell$ than the average extracted work $\langle W\rangle\sim \ln \ell$. 

As discussed above, the  proper operation of a continuous Maxwell demon relies on the assumption that $\delta s_{\rm meas}>0$ can be made arbitrary small.  Contrarily to the dissipation associated with memory erasure,  there is currently no known physical principle that limits the thermodynamic cost of measurements~\cite{leff1990maxwell}.  This is because the process of storing (erasing)  information  on a memory device is a physical  operation that increases (decreases) the  entropy of the device, while  the operation of measurement does not have a clear connection with entropy.      Nevertheless, physicists have investigated the cost of measurements with models of  external protocols.  Brillouin~\cite{brillouin1951maxwell} demonstrated that light cannot be used for this purpose, which led Bennett to develop a mechanical protocol to measure the position of the particle in Szilard's engine~\cite{bennett1987demons}.   This protocol is quasi-statically slow,  and is  thus not compatible with the high frequency measurements of continuous demons ($\delta t\rightarrow0$).    Therefore, it seems  reasonable to  associate a   nonzero, albeit small,  cost $\delta s_{\rm meas}$ to each  single measurement of a continuous demon.  According to  (\ref{sec:meas}) we can  then expect a significant measurement cost in the limit of continuous monitoring.

\subsection{Direct measurements} 
We discuss the measurement cost of setups that directly measure  first-passage times $T$.   As discussed in Sec.~\ref{sec:7},    a first-passage time detector is a device that sends out a signal as soon as the particle reaches the end of the bounded interval.

The measurements  in the setup of Fig.~\ref{fig:exampleMeas} have a thermodynamic cost, as the current in the circuit dissipates heat.   However, unlike continuous Maxwell demons,  this cost is independent of $\ell$, the size of the system.     Hence, considering a large enough system,  the cost of measurement can be made  negligible compared with the extracted work.     This follows from the fact that in this setup  first-passage times are measured with  a single measurement.  We leave a more detailed study of this setup for future work.   
 
\section{Discussion}\label{sec:9}

We have drawn  attention to  a  class of perpetuum mobile that use external protocols with a randomly timed step.   Several recent papers have studied such protocols, see Refs.~\cite{neri2019integral, neri2020second, hao,manzano2021thermodynamics, gambling2, gambling3,ribezzi2019large, ribezzi2019workv2, garrahan2023generalized, schmitt2023information, archambault2024first,tohme2024gambling}, demonstrating that the  limits of thermodynamics can seemingly be overcome.

In the present paper, we have resolved these thermodynamic puzzles.  The resolution, in a nutshell,  is that  temporal information must be treated on an equal footing with spatial information. Specifically, we have shown that the second law of thermodynamics is upheld once the random timing of events is recorded in a memory, as recording information on a tape increases the system entropy of the tape.     This explanation is analogous to the way the second law of thermodynamics is resolved in  Szilard’s engine~\cite{leff1990maxwell,feynman2018feynman}, the sole difference being in the nature of the random variables that is stored in the memory.   

Consequently, using random times in external protocols incurs a thermodynamic cost that is  informational in nature. This additional cost must also be accounted for in protocols not typically associated with Maxwell demons. A pertinent example is stochastic resetting, a process that maintains a nonequilibrium steady state by returning a diffusing particle to the origin at a random time~\cite{evans2011diffusion,evans2020stochastic}. This process requires an input of work, as energy must be expended to bring the particle back to the origin~\cite{olsen2024thermodynamic}. However, there is an additional cost related to storing the random time $T$ in memory. Neglecting this cost can lead to apparent violations of the second law of thermodynamics.

To  illustrate the  consequences of temporal information on nonequilibrium thermodynamics, we have introduced a  non-autonomous Maxwell demon that directly measures the random time $T$, and which  we called the sleeping demon.  
We have  developed setups that measure first-passage times directly,  unlike the conventional approach based on continuous monitoring  that is typically used in the literature~\cite{ schmitt2023information, ribezzi2019large, ribezzi2019workv2, garrahan2023generalized, archambault2024first}.     In the direct approach, the cost of measurement is fixed and independent of the size of the system, while for  continuous measurement the cost of measurement grows with the system size.

Sleeping demons need a clock in order to measure the time $T$ at the moment when they are "awake".  The requirement of a clock is not a specific feature of the sleeping demon, as clocks are  generally required for the operation of    non-autonomous Maxwell demons; even Maxwell's demon in  Szilard's engine needs a clock to operate.   The requirement of a clock  is 
   evident from the dependency of  $\lambda(t)$ on time $t$.  Thus, if a non-autonomous demon evaluates $\lambda$ it must know the time $t$.    Clocks are also required in certain measurement setups, such as in continuous measurements, as these setups measure the state of a system at evenly spaced time intervals. Thus the requirement of a clock in the measurement process is not specific to the sleeping demon model.  As the clock  operates at the same temperature as the system, maintaining its precision incurs a thermodynamic cost~\cite{barato2016cost,milburn2020thermodynamics, prech2024optimal,macieszczak2024ultimate,wadhia2025entropic}.  However, because the clock can serve a large number of parallel experiments, this cost does not scale with system size.  This contrasts with the cost of resetting the  memory that stores $T$, which does increase with system size.

Szilard engines have been implemented in  electronic devices, such as single electron boxes~\cite{koski2014experimental}, which raises the question of how to describe temporal information in quantum mechanics.     We expect some complications  due to the quantum Zeno effect~\cite{misra1977zeno,koshino2005quantum,dubey2021quantum,das2022quantum}.   There is also a possibility to define first-passage times in Bohmian mechanics~\cite{Nitta, str}, which may yield an alternative perspective on temporal information in quantum mechanics.    

Temporal information  is also relevant in  models of computing, such as those based on  finite-state machines~\cite{manzano2024thermodynamics}, and hence  the cost of temporal information should be considered in the thermodynamics of computation~\cite{bennett1982thermodynamics}.

\acknowledgements{The author thanks  K. S. Olsen for pointing out a few relevant references.  The author thanks  S. Bo and   M. Mitchison  for insightful discussions.}

\appendix

\section{Derivation of the bound Eq.~(\ref{eq:WBound2})}\label{app:A}

In this appendix we derive the inequality 
 \begin{equation}
\langle \exp\left(-\beta[ W(T+\tau)-\Delta f] +\ln p_T(T)\right) \rangle \leq 1.   \label{eq:WBound}
\end{equation}
  Applying Jensen's inequality to (\ref{eq:WBound}), we recover (\ref{eq:WBound2}).

  Note that it is possible to refine the result (\ref{eq:WBound}) into  an equality following   the approach in Ref.~\cite{ashida2014general}.   However, for our  purpose of deriving the second law for randomly timed protocols the  result (\ref{eq:WBound}) suffices.

 We derive the bound (\ref{eq:WBound}) for Markov chains in discrete time, even though the results also apply in continuous time.   Our derivations are based on  the framework of error-free non-autonomous Maxwell demons with a path-dependent protocol $\lambda(t;X^t_0)$, see Refs.~\cite{Sag2010, Sag2012, sagawa2013information}.    For the analysis below,   we require that $T$ is a stopping time (i.e., $T$ is independent of the values of $X$ after $T$).     

Let us consider a Markov chain $X(t)\in \mathcal{X}$ with $t\in \mathbb{N}$.   We denote the  transition probabilities by 
\begin{eqnarray}
\lefteqn{\mathbf{q}_{xy}(\lambda_s(t-1))}&& \nonumber\\ 
&& := {\rm Prob}\left\{X(t)= y| X(t-1)=x;\lambda_s(t-1)\right\},
\end{eqnarray}
where 
\begin{equation}
\lambda_s(t) = \left\{\begin{array}{ccc}  \lambda^{(1)}(t)&& t \in [0,s-1], \\   \lambda^{(2)}(t-s) &&  t\in [s,s+\tau-1],\\   \lambda_+&&  t\in [s+\tau,\infty),\end{array}\right.
\end{equation}
denotes an external time-dependent parameter, and ${\rm Prob}\left\{\cdot | \cdot ;\lambda_s(t-1)\right\}$ is a conditional probability that depends on $\lambda_s(t-1)$.     For now,  the time     $s\in\mathbb{N}$  when the protocol changes from $\lambda^{(1)}$ to $\lambda^{(2)}$ is  deterministic.   We denote the probability distribution of the initial configuration by $p_{\rm init}(x)$.

The probability that a sequence  $X^{s+\tau}_0= (X(0), X(1),\ldots, X(s+\tau))$ equals $x^{s+\tau}_0= (x(0),x(1),\ldots, x(s+\tau))$ is thus 
\begin{eqnarray}
\lefteqn{{\rm Prob}\left\{X^{s+\tau}_0 = x^{s+\tau}_0;\lambda^{s+\tau-1}_{0;s}\right\}} && 
\nonumber\\ 
&&= p_{\rm init}(x(0))\mathbf{q}_{x(0)x(1)}(\lambda_s(0)) \mathbf{q}_{x(1)x(2)}(\lambda_s(1))
\nonumber\\ 
&& 
\ldots \mathbf{q}_{x(s+\tau-1)x(s+\tau)}(\lambda_s(s+\tau-1)),
\end{eqnarray}
where $\lambda^{s+\tau-1}_{0;s} = (\lambda_s(0),\lambda_s(1),\ldots, \lambda_s(s+\tau-1))$. Let us  introduce the notation
\begin{eqnarray}
q(x^{s+\tau}_0;\lambda^{s+\tau-1}_{0;s}):= {\rm Prob}\left\{X^{s+\tau}_{0} = x_0^{s+\tau};\lambda^{s+\tau-1}_{0;s}\right\} \nonumber
\end{eqnarray} 
for  path probabilities.  Evaluating  
the path probability $q$  at  a stopping time, $s=T$,  and on a random trajectory, $x_{0}^{T+\tau}=X_0^{T+\tau}$, we obtain     
\begin{eqnarray}
\lefteqn{q(X_0^{T+\tau};\lambda^{T+\tau-1}_{0;T})}&& \nonumber\\ 
&&= p_{\rm init}(X(0)) \mathbf{q}_{X(0)X(1)}(\lambda_T(0)) \mathbf{q}_{X(1)X(2)}(\lambda_T(1))
\nonumber\\ 
&& 
\ldots \mathbf{q}_{X(T+\tau-1)X(T+\tau)}(\lambda_T(T+\tau-1)).
\end{eqnarray} 
Analogously, the time-reversed path probability is given by 
\begin{eqnarray}
\lefteqn{\tilde{q}(x_{0}^{s+\tau};\lambda^{s+\tau-1}_{0;s})} && \nonumber\\  
&&= \tilde{p}_{\rm init}(x(s+\tau)) \mathbf{q}_{x(s+\tau)x(s+\tau-1)}(\lambda_s(s+\tau-1))
\nonumber\\ 
&& 
\mathbf{q}_{x(s+\tau-1)x(s+\tau-2)}(\lambda_s(s+\tau-2)) \ldots \mathbf{q}_{x(1)x(0)}(\lambda_s(0)),\nonumber
\end{eqnarray} 
where $\tilde{p}_{\rm init}$ is the distribution of $X(0)$ in the time-reversed dynamics.    The ratio of  $\tilde{q}$ and $q$ takes the form 
\begin{eqnarray}
\lefteqn{\frac{\tilde{q}(X_0^{T+\tau};\lambda_{0;T}^{T+\tau-1})}{q(X_0^{T+\tau};\lambda_{0;T}^{T+\tau-1})}} && \nonumber \\ 
&=&  \frac{\tilde{p}_{\rm init}(X(T+\tau))}{p_{\rm init}(X(0))}\frac{\mathbf{q}_{X(1)X(0)}(\lambda_T(0))}{\mathbf{q}_{X(0)X(1)}(\lambda_T(0))}\frac{\mathbf{q}_{X(2)X(1)}(\lambda_T(1))}{\mathbf{q}_{X(1)X(2)}(\lambda_T(1))}
\nonumber\\ 
&& \cdots  \frac{\mathbf{q}_{X(T+\tau)X(T+\tau-1)}(\lambda_T(T+\tau-1))}{\mathbf{q}_{X(T+\tau-1)X(T+\tau)}(\lambda_T(T+\tau-1))}. \label{eq:ratio}
\end{eqnarray} 

Stochastic thermodynamics links the theory of Markov processes with nonequilibrium thermodynamics  through the principle of local detailed balance~\cite{bergmann1955new,maes2021local} 
\begin{equation}
\frac{\mathbf{q}_{xy}(\lambda)}{\mathbf{q}_{yx}(\lambda) }= \exp\left(S^{\rm env}_{xy}(\lambda)\right), \label{eq:princDetBal}
\end{equation}
with $S^{\rm env}_{xy}(\lambda)$ the entropy change in the environment during the transition from $x$ to $y$.   Using the formula (\ref{eq:princDetBal}) in Eq.~(\ref{eq:ratio})  for the ratio, we  obtain 
\begin{eqnarray}
\frac{\tilde{q}(X_0^{T+\tau};\lambda_{0;T}^{T+\tau-1})}{q(X_0^{T+\tau};\lambda_{0;T}^{T+\tau-1})}
&=& e^{-S_{\rm env}(T+\tau)}\frac{\tilde{p}_{\rm init}(X(T+\tau))}{p_{\rm init}(X(0))}\nonumber
\end{eqnarray} 
with $S_{\rm env}(T+\tau)$  the total  change in the entropy in the environment along the trajectory $X_0^{T+\tau}$.
If we consider an   isothermal system, as in the main text, then  
$S_{\rm env}(T+\tau)= \beta Q(T+\tau)$ with $Q(T+\tau)$ the heat released into the thermal reservoir.  If, in addition,   we set the initial conditions  
\begin{equation}
p_{\rm init}(x) = \exp\left(-\beta [u(x;\lambda_-)-f_-]\right)
\end{equation}
and 
\begin{equation}
\tilde{p}_{\rm init}(x) =  \exp\left(-\beta [u(x;\lambda_+)-f_+]\right),
\end{equation}
and we use the first law of thermodynamics 
\begin{equation}
u(X(T+\tau);\lambda_+)-u(X(0);\lambda_-) =  W(T+\tau) -  Q(T+\tau),  \label{eq:first}
\end{equation}
we get that 
\begin{eqnarray}
\frac{\tilde{q}(X_0^{T+\tau};\lambda^{T+\tau-1}_{0;T})}{q(X_0^{T+\tau};\lambda^{T+\tau-1}_{0;T})}  
&=& \exp\left(-\beta( W(T+\tau)-\Delta f)\right) \nonumber\\ \label{eq:ratiov2} , 
\end{eqnarray} 
with $\Delta f = f_+-f_-$; notice that in (\ref{eq:first}) $W$ is the work done on the system, while $Q$ is the heat released into the environment, and hence the sign difference.
Multiplying (\ref{eq:ratiov2}) by $p_T(T)$ and  taking an ensemble average over all trajectories $x^{\infty}_0$ weighted by  their path probability $q$, we obtain 
\begin{eqnarray}
\lefteqn{\langle \exp\left(-\beta( W-\Delta f) +\ln p_T(T)\right) \rangle } && \nonumber\\  
&=& \int_{\mathcal{X}^{\infty}} {\rm d} x^{\infty}_0q(x^{\infty}_0;\lambda^{\infty}_0) p_T(T(x^{\infty}_0)) \frac{\tilde{q}(x_0^{T+\tau};\lambda^{T+\tau-1}_{0;T})}{q(x_0^{T+\tau};\lambda^{T+\tau-1}_{0;T})}\nonumber\\
&=& \sum^{\infty}_{t=0}  p_T(t) \int_{\mathcal{X}^{t+\tau}} {\rm d} x^{t+\tau}_0 \: \delta_{t,T(x^{t+\tau}_0)}
\nonumber\\ 
&& \times
 \:  \tilde{q}(x_0^{t+\tau};\lambda^{t+\tau-1}_{0;t})  \int dx^{\infty}_{t+\tau+1}\: q(x^{\infty}_{t+\tau+1}|x(t+\tau);\lambda_+)\nonumber\\  
 &\leq&  \sum^{\infty}_{t=0}  p_T(t) \int_{\mathcal{X}^{t+\tau}} {\rm d} x^{t+\tau}_0  \tilde{q}( x^{t+\tau}_0;\lambda^{t+\tau-1}_{0;t})
 \nonumber\\
&= & 1, \label{eq:int}
\end{eqnarray}
which is the bound (\ref{eq:WBound}) that we were meant to derive.  Notice that in the second equality we introduced $\sum^{\infty}_{t=0}\delta_{t,T(x^{\infty}_0)}=1$,   we have used that $T$ is a stopping time to write $T(x^{\infty}_0) = T(x^{T+\tau}_0)$, and we have used the strong Markov property to write  $q(x^{\infty}_0,\lambda^{\infty}_0) = q(x^{T+\tau}_0;\lambda^{T+\tau-1}_0)q(x^{\infty}_{T+\tau+1}|x(T+\tau);\lambda_+) $~\cite{bremaud2001markov}.

\section{Entropy $H(T)$ for the random-time Szilard engine in discrete time and for $\ell=2$}\label{App:B}
We consider the  discrete, random-time Szilard engine  of Sec.~\ref{sec:5} for  $\ell=2$ and without measurement error.  In this case,   the distribution of the initiation time of the protocol is given by 
\begin{equation}
p_T(t) = \frac{1}{2} \delta_{t,0} + \frac{p_0 }{4}\sum^{\infty}_{j=1}  (1-p_0/2)^{j-1} \: \delta_{t,j}, 
\end{equation}
and its entropy is 
\begin{equation}
H(T) = -\frac{1}{2}\ln \frac{p_0}{8} -\frac{(1-p_0/2)}{p_0} \ln (1-p_0/2).  
\end{equation}
For $p_0=1$ this yields $H(T) = \ln \ell^2$, and this value is  annotated by a star in Figure~\ref{fig:plotEntropy}.

\section{Random-time Szilard engine in continuous time}\label{app:C}
We analyse the continuous version of the random-time Szilard engine, as introduced in Sec.~\ref{sec:6}.   As in the main text, we assume the demon has a perfect random-time detector at its disposal that measures $T$ without error, and  we also assume that the memory of the demon is  finite.

\subsection{Model for the working substance}
Let $X(t)\in [0,\ell]$ denote the position of a Brownian particle at time $t\in\mathbb{R}^+$ on a one-dimensional bounded domain of width $\ell$.    The evolution in time  of the position of the Brownian particle is described by 
\begin{equation}
\frac{{\rm d}X}{{\rm d}t} = \sqrt{2D} \frac{{\rm d}Z}{{\rm d}t}, 
\end{equation}
where $D$ is the diffusivity constant,  $Z(t)$ is a  standard Wiener process, and we implement reflecting boundary conditions at  $X(t)=0$ and $X(t)=\ell$.   We assume that at the initial time $t=0$, the particle is equilibrated, so that it is uniformly distributed over the domain $[0,\ell]$.

\subsection{Probability distribution of the first-passage time $T_{\rm fp}$}
The detector alerts the demon at the first-passage time  $T_{\rm fp}$ when the  Brownian particle hits the right boundary, see Eq.~(\ref{eq:Tfp}).   The first-passage time distribution $p_{T_{\rm fp}}$ can be obtained from the formula 
\begin{eqnarray}
p_{T_{\rm fp}}(t)  
&=& -\frac{1}{\ell}\int^{\ell}_0{\rm d}x_0 \: \partial_t \int^{\ell}_0 {\rm d}x\: G(x,t|x_0),  \label{eq:pT} 
\end{eqnarray}
where the propagator $G(x,t|x_0)$  is the probability of finding the Brownian particle at $X(t)=x$ if $X(0)=x_0$ (with an absorbing boundary condition at $x=\ell$).     The propagator solves~\cite{redner2001guide} 
\begin{equation}
\frac{\partial G(x,t|x_0)}{\partial t} = D\frac{\partial^2 G(x,t|x_0)}{(\partial x)^2}  \label{eq:G}
\end{equation}
with a reflecting  boundary condition at $x=0$,  
\begin{equation}
\left.\frac{\partial G(x,t|x_0)}{\partial x} \right|_{x=0} = 0,
\end{equation}
an absorbing boundary condition at $x=\ell$,
\begin{equation}
G(\ell,t|x_0) = 0,   
\end{equation}
and the initial condition, 
\begin{equation}
G(x,0|x_0) = \delta(x-x_0). \label{eq:initG}
\end{equation}

The solution to the Eqs.~(\ref{eq:G})-(\ref{eq:initG}) is  given by (see for example Ref.~\cite{huang2024extremal})
\begin{eqnarray}
G(x,t|x_0) &=& \frac{2}{\ell} \sum^{\infty}_{n=0} \cos \left(\frac{(n+1/2)\pi x_0}{\ell}\right) 
\nonumber\\ 
&& \times \cos \left(\frac{(n+1/2)\pi x}{\ell}\right) 
\nonumber\\ 
&& \times
\exp\left(-\frac{(n+1/2)^2\pi^2 Dt}{\ell^2}\right). \label{eq:GSol}
\end{eqnarray}
Substitution of (\ref{eq:GSol}) in Eq.~(\ref{eq:pT}),  taking the derivative with respect to time,  and evaluating the two integrals, we obtain 
\begin{equation}
p_{T_{\rm fp}}(t) = \frac{2D}{\ell^2} \sum^{\infty}_{n=0}      \exp\left(-\frac{(n+1/2)^2\pi^2 Dt}{\ell^2}\right).  \label{eq:pfptf}
\end{equation}
Notice that $p_{T_{\rm fp}}$ is normalised   and that $T_{\rm fp}$ scales as $\ell^2/D$ (see Appendix~\ref{sec:moments}).

%Hence, for large value of $\ell$, the ratio $\langle T\rangle/\sqrt{\langle T^2\rangle - \langle T\rangle^2}$ does not converge to zero, end hence there is no large deviation principle. 

%So let us therefore introduce a new variable $s = t/\ell^2$ so that 
%\begin{equation}
%q(s) =  p_T(s \ell^2) \ell^2 =  2D \sum^{\infty}_{n=0}      \exp\left(-(n+1/2)^2\pi^2 Ds\right)
%\end{equation}
%The times are $n\theta/\ell^2 $.

\subsection{The entropy $H(T)$ of the  random control time $T$}
The random time $T$ when the demon starts its protocol 
is defined by Eq.~(\ref{eq:TContDef})
where $\theta>0$ is a parameter denoting the time resolution of the control time.    
From the definition of $T$ in terms of $T_{\rm fp}$ it follows that the   probability mass function $p_T(t)$ of $T$ is given by 
\begin{eqnarray}
p_{T}(t) 
&=& \int^{(t+1)\theta}_{t\theta}  {\rm d}u \:  p_{T_{\rm fp}}(u) .
\end{eqnarray}
Using the  expression (\ref{eq:pfptf})  for  $p_{T_{\rm fp}}$ we get 
\begin{eqnarray}
p_{T}(t) 
&=& \frac{2 }{\pi^2}   \sum^{\infty}_{n=0}   \frac{1}{(n+1/2)^2} \exp\left(-t(n+1/2)^2\pi^2   \zeta \right)  
\nonumber\\ 
&& \times 
\left[1- \exp\left(-(n+1/2)^2\pi^2  \zeta \right)\right],  \label{eq:pTm}
\end{eqnarray}
where  $\zeta$ is the ratio 
  between the average (squared) distance $D\theta$ travelled by a Brownian particle  in a time interval $\theta$ and the squared interval width $\ell^2$ [see Eq.~(\ref{eq:zetaDef})]. 
 
\begin{figure}[t!]
\centering
{\includegraphics[width=0.4\textwidth]{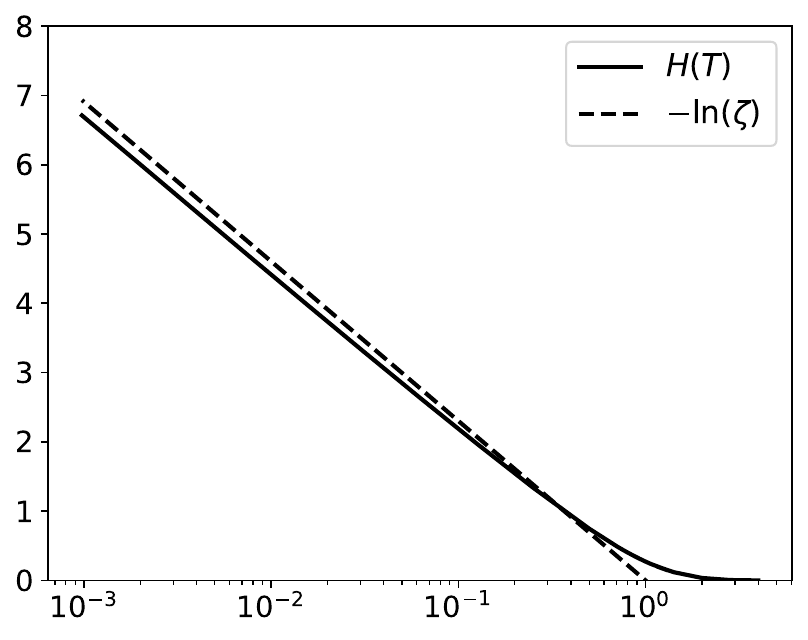}}
 \put(-100,-2){$\zeta$}
\caption{Entropy $H(T)$ of the distribution $p_T(t)$ given by Eq.~(\ref{eq:pTm}) as a function of $\zeta$.   The dashed line denotes $-\ln(\zeta)$ for comparison.  }\label{fig:plotEntropyV2} 
\end{figure}

 The entropy  $H(T)$ of $T$, as  defined by Eq.~(\ref{eq:entropyDef}),  is thus  a function of the parameter $\zeta$, and in  Fig.~\ref{fig:plotEntropyV2}  we plot the entropy as a function of $\zeta$.      Note that although $H(T)$ is a complicated function of $\zeta$ given by a double series, it  is well approximated by $-\ln \zeta $  for values of $\zeta<0.1$.  
 
\subsection{Average work $\langle W\rangle$}  \label{sec:WorkFig}
\begin{figure}[t!]
\hspace*{-1cm} 
{\includegraphics[width=0.6\textwidth]{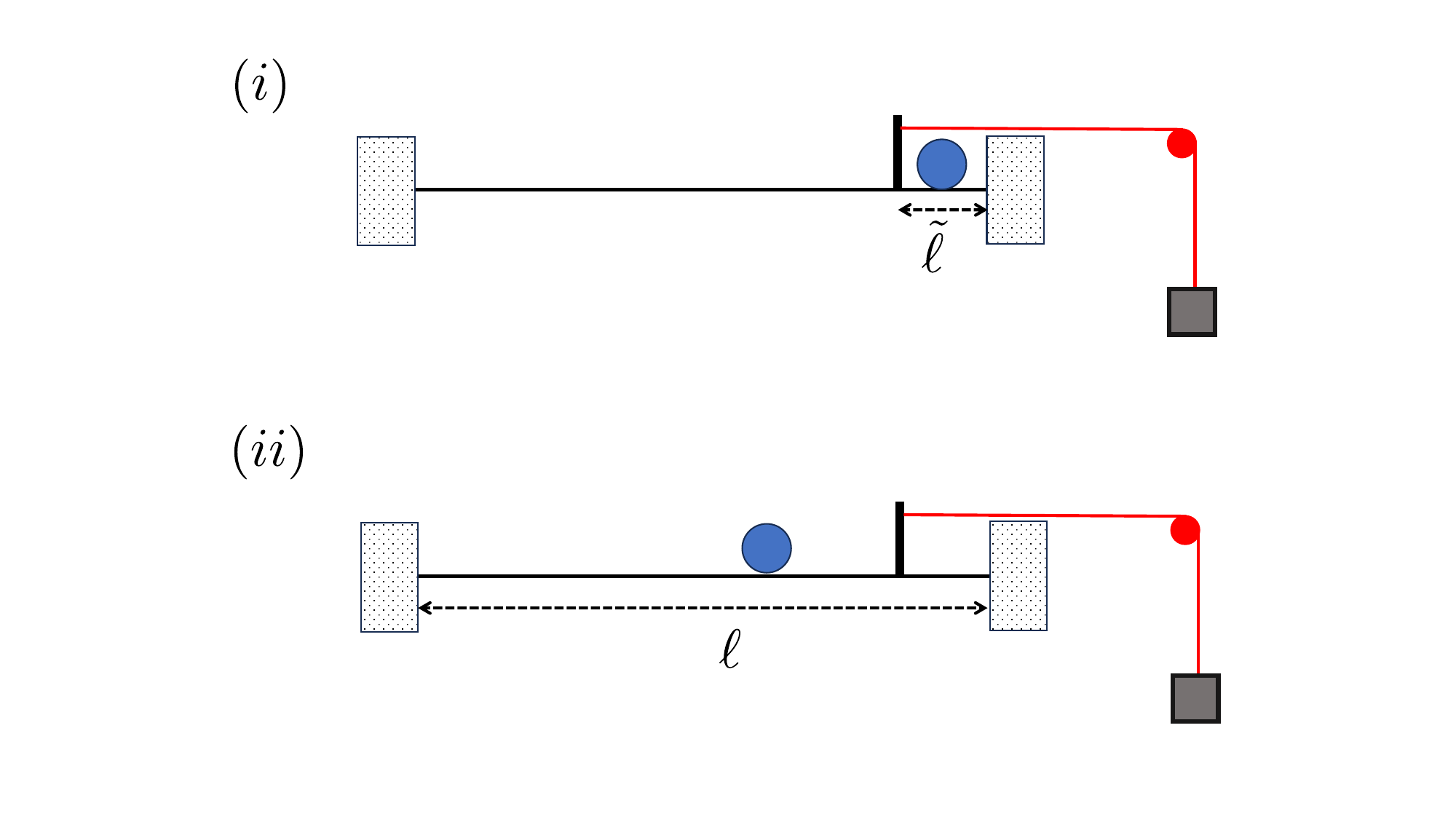}}
\caption{Case (i): $X(T)\in [\ell-\tilde{\ell},\ell]$, so that  the particle does  an amount $-\beta^{-1} \ln [(\ell-\tilde{\ell})/\tilde{\ell}] $ of   work on the external load. Case (ii):  $X(T)\in [0,\ell-\tilde{\ell}]$, so that  the load does  an amount $\beta^{-1} \ln (\ell-\tilde{\ell})/\tilde{\ell}$ of   work on the system.  %The star indicates the maximum of $\langle W(T+\tau)\rangle$ and the corresponding value of $p_+$. 
}\label{fig:maxwell2} 
\end{figure}

The average work done by the load on the particle is given by Eq.~(\ref{eq:Waverage}), with $p_+$ the probability that at time $T$ the particle is located right of the moveable wall.    The first term is negative and the second term is positive, as work is done by the  system on the external load  if $X(T)\geq \ell-\tilde{\ell}$, while work is done by the external load on the system if $X(T)\leq \ell-\tilde{\ell}$.    This is because the sleeping demon attaches the load to the right-hand side of the moveable wall for every realisation of the process (see Fig.~\ref{fig:maxwell2} for an illustration of both cases).

\begin{figure}[t!]
\centering
{\includegraphics[width=0.4\textwidth]{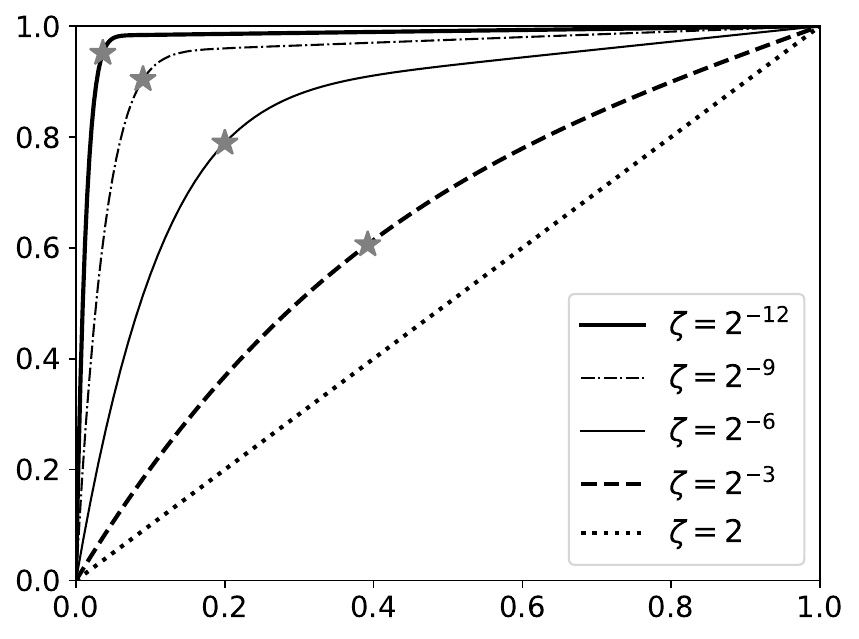}}
 \put(-215,70){$p_+$}
 \put(-100,-10){$\tilde{\ell}/\ell$}
 \hspace{1.5cm}
{\includegraphics[width=0.4\textwidth]{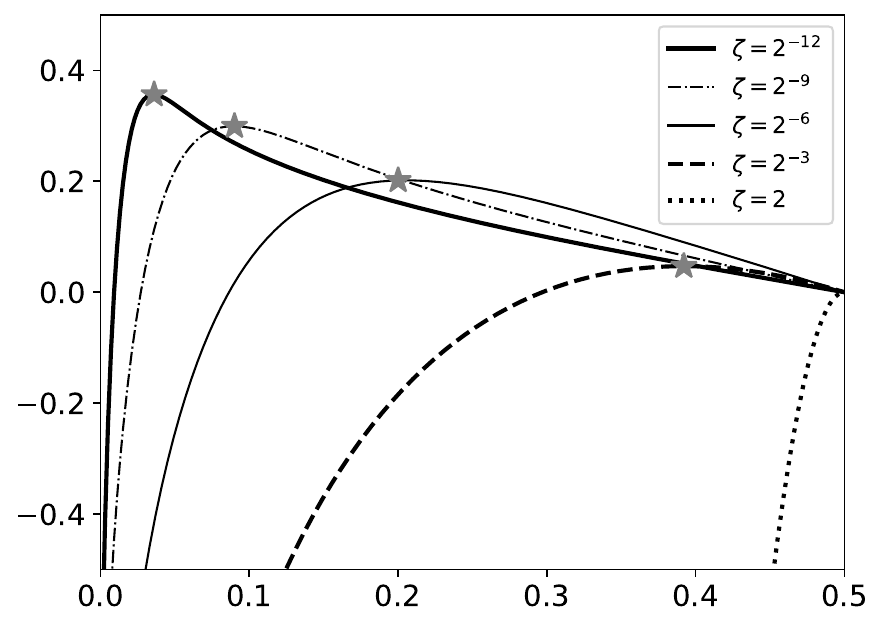}}
 \put(-245,70){$-\frac{\beta\langle W(T+\tau)\rangle}{ H(T)}$}
 \put(-100,-10){$\tilde{\ell}/\ell$}
\caption{Extended version of Fig.~\ref{fig:work} in the main text. Top Panel: The probability $p_+$ that $X(T)\geq \ell-\tilde{\ell}$ as a function of $\tilde{\ell}/\ell$ for different values of $\zeta = D\theta/\ell^2$ as indicated in the legend; the plot function is given by  Eq.~(\ref{eq:pP}).   Bottom Panel: The average work performed on the particle as a function of   $\tilde{\ell}/\ell$ (extended version of Fig.~\ref{fig:work} in the main text); for $\tilde{\ell}>\ell/2$ the average work is positive (not shown).  The star indicates the points on the curves evaluated at the optimal placement of the moveable wall, viz.,  $\tilde{\ell}/\ell = r^\ast$ with $r^\ast$ the value of the ratio at which $-\beta \langle W\rangle/H(T)$ attains its maximum.  }\label{fig:pPlus} 
\end{figure}

   As we show in Appendix.~\ref{sec:prob},  the  probability $p_+$  takes the form 
\begin{eqnarray}
\lefteqn{ p_{+} =\frac{\tilde{\ell}}{\ell}+ \frac{4}{\pi^3}\sum^{\infty}_{k=0}\frac{\exp\left((k+1/2)^2\pi^2\zeta\right)}{\exp\left((k+1/2)^2\pi^2\zeta\right)-1}}&& 
 \nonumber\\ 
&&\times  \sum^{\infty}_{n=1} \frac{\sin(n\pi \tilde{\ell}/\ell)}{n} \frac{\exp\left(-n^2 \pi^2\zeta \right)-\exp\left(-(k+1/2)^2 \pi^2\zeta \right)}{(1/2+k)^2-n^2}.
\nonumber\\ \label{eq:pP}
\end{eqnarray}
Note  that  $p_+$ is a function of  two parameters, viz., the ratios $\zeta = D\theta/\ell^2$ and $\tilde{\ell}/\ell$.    In the Top Panel of Fig.~\ref{fig:pPlus} we plot $p_+$ as a function of $\tilde{\ell}/\ell$ for several values of $\zeta$.  Even though the formal expression (\ref{eq:pP}) is a fairly complicated double series,   Fig.~\ref{fig:pPlus} shows that $p_+$ is a rather simple  monotonically  increasing function of $\tilde{\ell}$ with  $p_+=0$ if $\tilde{\ell}=0$ and $p_+=1$ if $\tilde{\ell}=\ell$.    In the limit $\zeta\rightarrow 0$, $p_+$ is well approximated by (see Top Panel of Fig.~\ref{fig:pPlus})
\begin{equation}
p_+ \approx  \left\{\begin{array}{ccc} 1 &{\rm if}&  \frac{\tilde{\ell}}{\ell} >   r^\ast ,\\  \frac{1}{r^\ast}\frac{\tilde{\ell}}{\ell} &{\rm if}&  \frac{\tilde{\ell}}{\ell} <   r^\ast , \end{array} \right.  \label{eq:pPRAST}
\end{equation}
where $r^\ast$ denotes the value of $\tilde{\ell}/\ell$   where the slope of $p_+$ changes from almost vertical to horizontal.

Using  the expression (\ref{eq:pP}) for $p_+$ in Eq.~(\ref{eq:Waverage}), we obtain  a formula for the average work $\langle W(T+\tau)\rangle$.    The    plot  of $-\beta \langle W\rangle/ H(T)$ in  Fig.~\ref{fig:work} of the Main Text is obtained from this formula together  with the Eq.~(\ref{eq:pTm}) that yields the Shannon entropy $H(T)$.  We provide a more detailed version of this figure in the Bottom Panel of Fig.~\ref{fig:pPlus}.   The stars in the Bottom Panel of Fig.~\ref{fig:pPlus} indicate the maximum value of $-\beta \langle W(T+\tau)\rangle/H(T)$, corresponding with the best  placement of the moveable wall.    We observe that the maximum increases as a function of $1/\zeta$, which raises the question of what is the maximum value of the ratio $-\beta \langle W(T+\tau)\rangle/H(T)$ in the limit $\zeta\rightarrow 0$? 

\begin{figure}[t!]
\centering
{\includegraphics[width=0.4\textwidth]{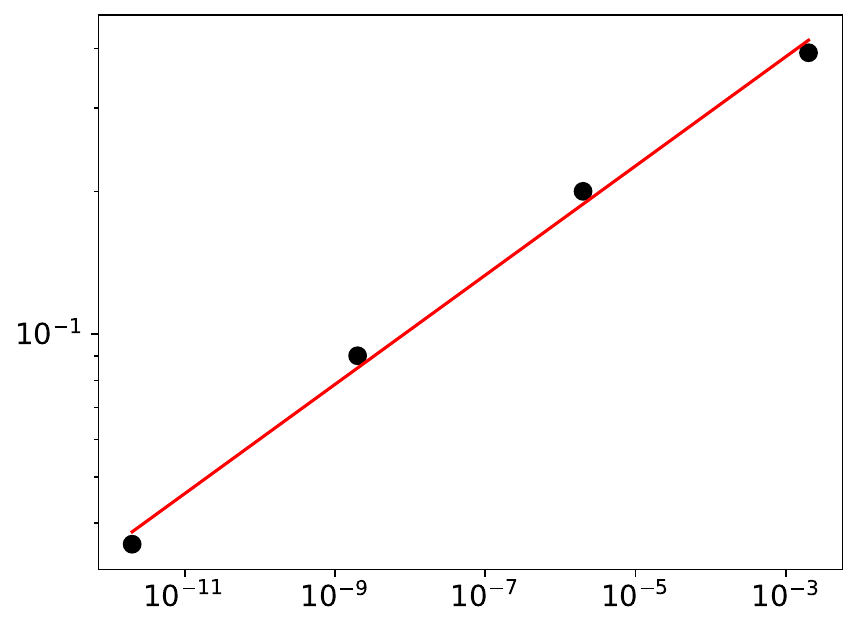}}
 \put(-215,70){$r^\ast$}
 \put(-90,-10){$\zeta$}
\caption{A plot of $r^\ast$, the optimal position to place the moveable wall [as defined in Eq.~(\ref{eq:rast})], as a function of  $\zeta=D\theta/\ell^2$.     The markers are obtained by numerically finding  the maximum of the curves in the Bottom Panel of Fig.~\ref{fig:pPlus}.   The solid line is a fit to the  power law function $\gamma \zeta^\alpha$; the  numerically obtained optimal fitting parameters are   $\gamma=0.92$ and $\alpha=0.38$.   }\label{fig:pPlus2} 
\end{figure}

To address this latter question, we define 
\begin{equation}
r^\ast = {\rm argmin}_{\tilde{\ell}/\ell} \frac{\beta \langle W(T+\tau)\rangle}{H(T)}, \label{eq:rast}
\end{equation}
as the value of the ratio $\tilde{\ell}/\ell$ when $-\beta \langle W(T+\tau)\rangle/H(T)$ attains its maximum.  
In the  Top Panel of Fig.~\ref{fig:pPlus} we plot $p_+(r^\ast)$ using the same star symbol as in the Bottom Panel of Fig.~\ref{fig:pPlus}.  This reveals that $r^\ast$, as defined in (\ref{eq:rast}), roughly equals the value of $\tilde{\ell}/\ell$ when the slope of $p_+$ changes from almost vertical to almost horizontal line (in the limit of small $\zeta$); this justifies why we have used the same symbol of $r^\ast$ in the Eqs.~(\ref{eq:pPRAST}) and (\ref{eq:rast}).    In Fig.~\ref{fig:pPlus2} we plot $r^\ast$ as a function of $\zeta$, and this reveals that 
\begin{equation}
r^\ast \approx \gamma \zeta^\alpha,
\end{equation}
with $\gamma=0.92$  and $\alpha=0.38$.   Thus, using Eq.~(\ref{eq:Waverage}) to evaluate  $\langle W\rangle$ for $\zeta\approx 0$ and $\tilde{\ell}/\ell = r^\ast$, and using that $p_+(r^\ast)\approx 1$ and  $r^\ast \approx\gamma \zeta^\alpha $, we get 
\begin{equation}
\beta \langle W(T+\tau )\rangle \approx  \alpha \ln \zeta   .   \label{eq:WT}
\end{equation}  
As  the entropy  $H(T)\approx -\ln (\zeta)$ in the limit $\zeta\approx 0$ (see Fig.~\ref{fig:plotEntropyV2}), we get in this limit that for $\tilde{\ell}/\ell = r^\ast$ the average work equals
\begin{equation}
-\beta \frac{\langle W(T+\tau )\rangle}{H(T)} \approx  \alpha   = 0.38 .    \label{eq:WT}
\end{equation}  
The  value $0.38$ for the maximum of the ratio $-\beta \frac{\langle W(T+\tau )\rangle}{H(T)}$ is  in good correspondence with the numerics in the Bottom Panel of Fig.~\ref{fig:pPlus}.  Indeed, the figure  shows that  the maximum amount of  extractable  work converges to $0.38 H(T)/\beta$ for $\zeta\rightarrow 0$.

%Considering that in the limit of $\tilde{\ell}\rightarrow 0$, 
%\begin{equation}
%\beta \langle W(T+\tau )\rangle =  -(1-2p_+) \ln \frac{\tilde{\ell}}{\ell} + O() =(1-2p_+) \ln r +  \frac{1}{2}(1-2p_+) \ln \zeta .   \label{eq:WT2}
%\end{equation} 

\section{Moments of the  distribution of $T_{\rm fp}$}  \label{sec:moments}
We determine  the moments of  the first-passage time $T_{\rm fp}$   with a  probability  distribution given by (\ref{eq:pfptf}).  

We rely on known results for   Hurwitz zeta functions defined by (see pages 19-22 in Ref.~\cite{magnus1966formulas}) 
\begin{equation}
\zeta(2k,x) = \sum^{\infty}_{n=0}(n+x)^{-2k}.  \label{eq:Hurwitz}
\end{equation}
As shown in Ref.~\cite{cvijovic1999values},   the Hurwitz zeta function satisfies
 \begin{equation}
 \zeta(2k,x) + \zeta(2k,1-x) = -\frac{\pi}{(2k-1)!}  \frac{{\rm d}^{2k-1}}{({\rm d}x)^{2k-1}} {\rm cot}(\pi x), \label{eq:zeta2n}
 \end{equation}
where $\cot(y) = 1/\tan(y)$ is the cotangent.  Setting $x=1/2$ in (\ref{eq:zeta2n}), we find that  
\begin{equation}
\zeta(2k,1/2) =  -\frac{\pi}{2(2k-1)!}\left.  \frac{{\rm d}^{2k-1}}{({\rm d}x)^{2k-1}} {\rm cot}(\pi x)\right|_{x=1/2} . \label{eq:zeta2nx}
\end{equation}
We will repeatedly use the formula (\ref{eq:zeta2nx}) at  different values of $k\in \mathbb{N}$.  

\subsection{Normalisation}\label{app:norm}
That $p_{T_{\rm fp}}$ is normalised  follows from  
\begin{equation}
\int^{\infty}_0 {\rm d}t \: p_{T_{\rm fp}}(t) =  \frac{2}{\pi^2} \sum^{\infty}_{n=0}\frac{1}{(n+1/2)^2} = \frac{2}{\pi^2}\zeta(2,1/2) .\label{eq:normFind}
\end{equation}
Using (\ref{eq:zeta2nx}) for $k=1$, we get 
\begin{equation}
2\zeta(2,1/2) = -\pi  \left.\frac{{\rm d}}{{\rm d}x}{\rm cot}(\pi x)\right|_{x=1/2} = \pi^2,
\end{equation}
which together with (\ref{eq:normFind}) implies the normalisation of $p_{T_{\rm fp}}(t)$.

\subsection{Mean first-passage time}\label{app:mean}
We show that  the mean first-passage time   is given by 
\begin{equation}
\int^{\infty}_0 {\rm d}t \:t\: p_{T_{\rm fp}}(t) =  \frac{2\ell^2}{\pi^4 D} \sum^{\infty}_{n=0} \frac{1}{(n+1/2)^4} = \frac{1}{3}\frac{\ell^2}{D}. \label{eq:meanTfP}
\end{equation}
Indeed, substituting (\ref{eq:pfptf}) in  the integral on the left-hand side of (\ref{eq:meanTfP}) and evaluating the integral in $t$ we obtain the series
\begin{equation}
\int^{\infty}_0 {\rm d}t \:t\: p_{T_{\rm fp}}(t) =  \frac{2\ell^2}{\pi^4 D} \sum^{\infty}_{n=0} \frac{1}{(n+1/2)^4}  = \frac{2 \ell^2}{\pi^4D}\zeta(4,1/2) ,\label{eq:meanFind}
\end{equation}
where in the last equality we have used the definition (\ref{eq:Hurwitz}) of the Hurwitz zeta function.
Using (\ref{eq:zeta2nx}) at $k=2$ we obtain $\zeta(4,1/2) = \pi^4/6$, and thus  (\ref{eq:meanFind}) implies (\ref{eq:meanTfP}).

\section{Probability $p_+$ that $X(T)\geq \ell-\tilde{\ell}$}  \label{sec:prob}
We derive the formula (\ref{eq:pP}) for  $p_+$. 
The probability $p_+$ can be expressed in terms of   the propagator $F(x,t|\ell)$ that gives the probability of $X(t)=x$ if $X(0)=\ell$: 
\begin{equation}
 p_{+} =\sum^{\infty}_{m=0}\int^{\theta}_{0} {\rm d}u\: p_{T_{\rm fp}}((m+1)\theta-u) \int^{\ell}_{\ell-\tilde{\ell}} {\rm d}x \: F(x,u|\ell). \label{eq:pPv3}
\end{equation} 

The propagator $F$ solves the equation  
\begin{equation}
\frac{\partial F(x,t|\ell)}{\partial t} = D\frac{\partial^2 F(x,t|\ell)}{(\partial x)^2}  
\end{equation}
with reflecting boundary conditions 
\begin{equation}
\left.\frac{\partial F(x,t|\ell)}{\partial x} \right|_{x=0} = 0 \quad {\rm and} \quad  \left.\frac{\partial F(x,t|\ell)}{\partial x} \right|_{x=\ell} = 0,
\end{equation}
and initial condition 
\begin{equation}
F(x,0|\ell) = \delta(x-\ell).  
\end{equation} 
The solution for $F$ is  given by~\cite{redner2001guide} 
\begin{eqnarray}
F(x,t|\ell) &=& \frac{1}{\ell}  
\nonumber\\ 
&& 
+ \frac{2}{\ell} \sum^{\infty}_{n=1}(-1)^n\cos\left(n\pi \frac{x}{\ell}\right) \exp\left(-\frac{n^2\pi^2 D t}{\ell^2}\right) . \nonumber\\\label{eq:FSol}
\end{eqnarray}

Substitution of (\ref{eq:FSol}) in (\ref{eq:pPv3}) gives 
\begin{eqnarray}
 p_{+} &=&\sum^{\infty}_{m=0}\int^{\theta}_{0} {\rm d}u\: p_{T_{\rm fp}}((m+1)\theta-u) \int^{\ell}_{\ell-\tilde{\ell}} {\rm d}x \: F(x,u|\ell) 
 \nonumber\\ 
 &=& \frac{\tilde{\ell}}{\ell}+  \frac{4D}{\ell^3} \sum^{\infty}_{k=0} \sum^{\infty}_{m=0}  \exp\left(-\frac{ m\theta(k+1/2)^2\pi^2 D}{\ell^2}\right)
 \nonumber\\ 
 && \times 
 \sum^{\infty}_{n=1}(-1)^n \int^{\ell}_{\ell-\tilde{\ell}} {\rm d}x \cos\left(n\pi \frac{x}{\ell}\right)
\nonumber\\ 
&& \times 
\int^{\theta}_{0} {\rm d}u \:   \exp\left(-\frac{(k+1/2)^2\pi^2 D (\theta-u)}{\ell^2}\right) \nonumber\\ 
&& \quad \quad \quad \quad  \times \exp\left(-\frac{n^2\pi^2 D u}{\ell^2}\right) .
 \label{eq:pP2}
\end{eqnarray} 
Evaluating the sums over $m$ and the integrals over $u$ and $x$ we obtain the expression (\ref{eq:pP}).

\bibliography{biblio}

 \end{document}